\documentclass[10pt]{bmc_article}

\usepackage{cite} 
\usepackage{url}  
\usepackage{ifthen}  
\usepackage{multicol}   
\usepackage[utf8]{inputenc} 
\usepackage{amssymb}
\usepackage{amsmath}
\usepackage{textcomp}
\usepackage{tipa}
\usepackage{wasysym}
\usepackage{graphicx}
\usepackage{bm}
\usepackage{revsymb}
\newboolean{publ}

\newenvironment{bmcformat}{\begin{raggedright}\baselineskip20pt\sloppy\setboolean{publ}{false}}{\end{raggedright}\baselineskip20pt\sloppy}

\begin{document}
\begin{bmcformat}


\title{Ge/Si(001) heterostructures with dense arrays of Ge \\ quantum dots: morphology, defects, photo-emf spectra \\ and terahertz conductivity
}
 

\author{Vladimir A Yuryev\correspondingauthor$^{1,2}$%
       \email{Vladimir A Yuryev\correspondingauthor 
- vyuryev@kapella.gpi.ru}%
\and
         Larisa V Arapkina$^{1}$
         \email{Larisa V Arapkina - arapkina@kapella.gpi.ru}%
\and
         Mikhail S Storozhevykh$^{1}$
         \email{Mikhail S Storozhevykh - storozhevykh@kapella.gpi.ru}%
\and
        Valery A Chapnin$^{1}$
         \email{Valery A Chapnin - chapnin@kapella.gpi.ru}%
\and
         Kirill V Chizh$^{1}$
         \email{Kirill V Chizh - chizh@kapella.gpi.ru}%
\and
         Oleg V Uvarov$^{1}$
         \email{Oleg V Uvarov - uvarov@kapella.gpi.ru}%
\and
         Victor P Kalinushkin$^{1,2}$
         \email{Victor P Kalinushkin - vkalin@kapella.gpi.ru}%
\and
         Elena S Zhukova$^{1,3}$
         \email{Elena S Zhukova - zhukovaelenka@gmail.com}%
\and
         Anatoly S Prokhorov$^{1,3}$
         \email{ Anatoly S Prokhorov - aspro@ran.gpi.ru}%
\and
         Igor E Spektor$^{1}~$
         \email{Igor E Spektor - spektor@ran.gpi.ru}%
   and
        Boris P Gorshunov$^{1,3}$
         \email{Boris P Gorshunov - gorshunov@ran.gpi.ru}%
      }


\address{%
    \iid(1)
A~M~Prokhorov General Physics Institute of RAS, 38 Vavilov Street, Moscow 119991, Russia\\
    \iid(2)Technopark of GPI RAS, 38 Vavilov Street, Moscow, 119991, Russia\\
    \iid(3)Moscow Institute of Physics and Technology, Institutsky Per. 9, Dolgoprudny, Moscow Region, 141700, Russia %
}%

\maketitle


\begin{abstract}
      


\paragraph*{Morphology and defects:}
Issues of Ge hut cluster array formation and growth at low temperatures on the Ge/Si(001) wetting layer are discussed on the basis of explorations performed by high resolution STM and {\it in-situ} RHEED. Dynamics of the RHEED patterns in the process of Ge hut array formation is investigated at low and high temperatures of Ge deposition. Different dynamics of RHEED patterns during the deposition of Ge atoms in different growth modes is observed, which reflects the difference in adatom mobility and their `condensation' fluxes from Ge 2D gas on the surface for different modes, which in turn control the nucleation rates and densities of Ge clusters. Data of HRTEM studies of multilayer Ge/Si heterostructures are presented with the focus on low-temperature formation of perfect films.
\paragraph*{Photo-emf spectroscopy:}
Heteroepitaxial Si {\textit{p--i--n}}-diodes with multilayer stacks of Ge/Si(001) quantum dot dense arrays built in intrinsic domains have been investigated and found to exhibit the photo-emf in a wide spectral range from 0.8 to 5\,$\mu$m. An effect of wide-band irradiation by infrared light on the photo-emf spectra has been observed. Photo-emf in different spectral ranges has been found to be differently affected by the wide-band irradiation. A significant increase in photo-emf is observed in the fundamental absorption range under the wide-band irradiation. The observed phenomena are explained in terms of positive and neutral charge states of the quantum dot layers and the Coulomb potential of the quantum dot ensemble. A new design of quantum dot infrared photodetectors is proposed.
\paragraph*{Terahertz spectroscopy:}
By using a coherent source spectrometer, first measurements of terahertz dynamical conductivity (absorptivity) spectra of Ge/Si(001) heterostructures were performed at frequencies ranged from 0.3 to 1.2 THz in the temperature interval from 300 to 5\,K. The effective dynamical conductivity of the heterostructures with Ge quantum dots has been discovered to be significantly higher than that of the structure with the same amount of bulk germanium (not organized in an array of quantum dots). The excess conductivity is not observed in the structures with the Ge coverage less than 8\,\AA. When a Ge/Si(001) sample is cooled down the conductivity of the heterostructure decreases.
\end{abstract}

\ifthenelse{\boolean{publ}}{\begin{multicols}{2}}{}


\section*{Introduction}

Artificial low-dimensional nano-sized objects, like quantum dots, quantum wires and quantum wells, as well as structures based on them, are promising systems for improvement of existing devices and for development of principally new devices for opto-, micro- and nano-electronics. Besides, the investigation of physical properties of such structures is also of fundamental importance. In both regards, amazing perspectives are provided when playing around with quantum dots that can be considered as artificial atoms with a controlled number of charge carriers that have a discrete energy spectrum \cite{Pchel_Review-TSF, Pchel_Review}. Arrays of a \textit{large} number of quantum dots including multilayer heterostructures make it possible to create artificial ``solids" whose properties can be controllably changed by varying the characteristics of constituent elements (``atoms") and/or the environment (semiconductor matrix). The rich set of exciting physical properties in this kind of systems originates from single-particle and collective interactions that depend on the number and mobility of carriers in quantum dots, Coulomb interaction between the carriers inside a quantum dot and in neighbouring quantum dots, charge coupling between neighbouring quantum dots, polaron and exciton effects, etc. Since characteristic energy scales of these interactions (distance between energy levels, Coulomb interaction between charges in quantum dots, one- and multiparticle exciton and polaron effects, plasmon excitations, etc.) are of order of several meV \cite{3-Colomb_interactions-Dvur,4-Drexler-InGaAs,5-Lipparini-far_infrared}, an appropriate experimental tool for their study is provided by optical spectroscopy in the far-infrared and terahertz bands.\pb 

To get access to the effects, one has to extend the operation range of the spectrometers to the corresponding frequency domain that is to the terahertz frequency band. Because of inaccessibility of this band, and especially of its lowest frequency part, below 1 THz (that is $\apprle 33$\,cm$^{-1}$), for standard infrared Fourier-transform spectrometers, correspondent data is presently missing in the literature. In this paper, we present the results of the first detailed measurements of the absolute dynamical (AC) conductivity of multilayer Ge/Si heterostructures with Ge quantum dots, at terahertz and sub-terahertz frequencies and in the temperature range from 5 to 300\,K.
\pb

In addition, for at least two tens of years, multilayer Ge/Si heterostructures with quantum dots have been candidates to the role of  photosensitive elements of monolithic IR arrays promising to replace and excel platinum silicide in this important brunch of the sensor technology \cite{Wang-properties,Wang-Cha,Dvur-IR-20mcm}. Unfortunately, to date achievements in this field have been less than modest. 
\pb

We believe that this state of affairs may be improved by rigorous investigation of formation, defects and other aspects of materials science of such structures, especially those which may affect device performance and reliability, focusing on identification of reasons of low quantum efficiency and detectivity, high dark current and tend to degrade with time as well as on search of ways to overcome these deficiences. New approaches to device architecture and design as well as to principles of functioning are also desirable.
\pb

This article reports our latest data on morphology and defects of Ge/Si heterostructures. 
On the basis of our recent results on the photo-emf in the Si {\textit{p--i--n}}-structures with Ge quantum dots, which are also reported in this article, we  propose  a new design of photovoltaic quantum dot infrared photodetectors.

\section*{Methods}

  \subsection*{Equipment and techniques}    

The Ge/Si samples were grown and characterized using an integrated ultrahigh vacuum
instrument \cite{classification,stm-rheed-EMRS,CMOS-compatible-EMRS,VCIAN2011} built on the basis of the Riber~SSC\,2
surface science center with the  EVA\,32 molecular-beam epitaxy (MBE) chamber  equipped with the RH20 reflection high-energy electron diffraction (RHEED) tool (Staib Instruments) and connected through a transfer line to the
 GPI-300 ultrahigh vacuum scanning tunnelling microscope (STM)
\cite{gpi300,STM_GPI-Proc,STM_calibration}. Sources with the electron beam evaporation were used for Ge or Si deposition. 
A Knudsen effusion cells was utilized if boron doping was applied for QDIP {\textit{p--i--n}}-structure  formation.
The pressure of about $5\times 10^{-9}$\,Torr was kept in the preliminary sample cleaning
(annealing) chamber. 
The MBE chamber was evacuated down to about $10^{-11}$\,Torr before processes; the pressure increased to nearly $2\times 10^{-9}$\,Torr at most during the Si substrate cleaning and $10^{-9}$\,Torr during Ge or Si deposition.   
The residual gas pressure did not exceed $10^{-10}$\,Torr in the STM chamber. 
Additional  details of the experimental instruments and process control can be found in Ref.\,\cite{VCIAN2011}.
\pb

RHEED measurements were carried out {\it in situ}, i.e., directly in the
MBE chamber during a process \cite{stm-rheed-EMRS}. 
STM images were obtained in the
constant tunnelling current 
mode at the room
temperature. The STM tip was zero-biased while the sample was
positively or negatively biased 
when scanned in empty-
or filled-states imaging mode.
Structural properties of the Ge/Si films were explored by using the Carl Zeiss Libra-200 FE HR HRTEM. 
\pb

The images were processed using the WSxM software \cite{WSxM}.\pb

For obtaining spectra of photo-electromotive force (photo-emf) a setup enabling sample illumination by two independent beams was used; one of the beams was a wide-band infrared (IR) radiation, generated by a tungsten bulb, passed through a filter of Si or Ge (bias lighting) and the other was a beam-chopper modulated narrow-band radiation cut from globar emission by an IR monochromator tunable in the range from 0.8 to 20 $\mu$m. The spectra were taken at the chopping frequency of 12.5 Hz at temperatures ranged from 300 to 70 K and a widely varied power of the bias lighting. 
\pb

The measurements of the terahertz dynamic conductivity and absorptivity of Ge/Si heterostructures at room and cryogenic  temperatures (down to 5 K) have been performed using the spectrometer based on backward-wave oscillators (BWO) as radiation sources. This advanced experimental technique will be described in detail below in a separate section.
\pb

\subsection*{Sample preparation procedures}  

\subsubsection*{Preparation of samples for STM and RHEED}

Initial samples for STM and RHEED studies were  $8\times 8$~mm$^{2}$ squares cut from the
specially treated commercial boron-doped    Czochralski-grown (CZ) Si$(100)$ wafers
($p$-type,  $\rho\,= 12~{\Omega}\,$cm).  After washing and chemical treatment
following the standard procedure described elsewhere \cite{cleaning_handbook}, which included washing in
ethanol, etching in the mixture of HNO$_3$ and HF and rinsing in the
deionized water \cite{VCIAN2011}, the silicon substrates were 
loaded into the airlock and transferred into the preliminary
annealing chamber where they were outgassed at the temperature of
around {565{\textcelsius}} for more than 6\,h. After that, the substrates were
moved for final treatment and Ge deposition into the MBE chamber 
where they were subjected to 
two-stages annealing during heating with stoppages at {600{\textcelsius}}
for 5\,min and at {800{\textcelsius}} for 3\,min \cite{classification,stm-rheed-EMRS}. The
final annealing at the temperature greater than {900{\textcelsius}} was
carried out for nearly 2.5\,min with the maximum temperature of
about {925{\textcelsius}} (1.5\,min). Then, the temperature was rapidly
lowered to about {750{\textcelsius}}. The rate of the further cooling was
around 0.4{\textcelsius}/s that corresponded to the `quenching' mode
applied in \cite{stm-rheed-EMRS}. The
surfaces of the silicon substrates were completely purified of the
oxide film as a result of this treatment \cite{our_Si(001)_en,phase_transition,stm-rheed-EMRS}.
\pb
 
Ge was deposited directly on the deoxidized Si(001) surface. The
deposition rate was varied from about $0.1$ to $0.15$\,\r{A}/s; the effective Ge film
thickness $(h_{\rm Ge})$ was varied from 3 to 18\,\r{A} for different
samples.  The substrate
temperature during Ge deposition $(T_{\rm gr})$ was {360{\textcelsius}}  for the low-temperature mode   and 600 or {650{\textcelsius}} for the  high-temperature mode.
The rate of the sample cooling down to the room temperature was approximately
0.4{\textcelsius}/s after the deposition.
\pb

\subsubsection*{Preparation of multilayer structures}

Ge/Si heterostructures with buried Ge layers were grown on CZ $p$-Si$(100)$:B wafers
($\rho\,= 12~{\Omega}\,$cm) washed and outgassed as described above. Deoxidized Si(001) surfaces were prepared by a  process allowed us to obtain clean substrate surfaces (this was verified by STM and RHEED) and perfect epitaxial interfaces with Si buffer layers (verified by HRTEM): the wafers were annealed at 800{\textcelsius} under Si flux of $\apprle 0.1$\,\AA/s until a total amount of the deposited Si, expressed in the units of the   Si film thickness indicated by the film thickness monitor, reached 30\,{\AA}; 2-minute stoppages of Si deposition  were made  first twice after every 5\,\r{A} and then twice after every 10\,\r{A}. 
\pb

Afterwards, a $\sim$\,100\,nm thick Si buffer was deposited on the prepared surface at the temperature of $\sim$\,650{\textcelsius}. Then, a single Ge layer or a multilayer Ge/Si structure was grown. A number of Ge layers in multilayer structures reached 15 but usually was 5; their effective thickness ($h_{\rm Ge}$), permanent for each sample, was varied from sample to sample in the range from 4 to 18\,\AA; the thickness of the Si spacers ($h_{\rm Si}$) was  $\sim$\,50\,nm. The Ge deposition temperature was $\sim$\,360{\textcelsius}, Si spacers were grown at $\sim$\,530{\textcelsius}. 
A heterostructure formed in such a way was capped by a  $\sim$\,100\,nm thick Si layer grown at $\sim$\,530{\textcelsius}. All layers were undoped.
\pb

The samples were quenched after the growth  at the rate of $\sim$\,0.4{\textcelsius}/s.
\pb

\subsubsection*{{Growth of \textit{p--i--n}}-structures}

 {\textit{p--i--n}}-structures were grown on commercial  phosphorus-doped CZ $n$-Si(100) substrates ($\rho = 0.1\,\Omega$\,cm). 
Si surfaces were prepared for structure deposition in the same way as  for the growth of multilayer structures. $i$-Si buffer domains of various thicknesses were grown on the clean surfaces at $\sim$\,650{\textcelsius}.  Then, a stacked structure of several periods of  quantum dot (QD) dense arrays separated by Si barriers was grown under the same conditions as the multilayer structures;  $h_{\rm Si}$ was widely varied in different structures reaching 50\,nm;  $h_{\rm Ge}$ always was 10\,\AA. A sufficiently thick undoped Si layer separated the stacked QD array from the Si:B cap doped during the growth, the both layers were grown at $\sim$\,530{\textcelsius}.
\pb

Figure~\ref{fig:p-i-n_Schematics} demonstrates two such structures (referred to as R\,163 and R\,166) which are  in the focus of this article. Their caps were doped to $5\times 10^{18}$ and $10^{19}$\,cm$^{-3}$ in the R\,163 and R\,166 samples, respectively. Buffer layer and barrier thicknesses were 99 and 8\,nm in the R\,163 structure and 1690 and 30\,nm in R\,166.
\pb

Mesas were formed on samples for photoelectric measurements. Ohmic contacts were formed by thermal deposition of aluminum.
\pb

\subsection*{Terahertz BWO-spectroscopy}  
 The BWO-spectrometers provide broad-band operation (frequencies $\nu$ ranging from 30 GHz to 2 THz), high frequency resolution ($\Delta \nu/\nu = 10^{-5}$), broad dynamic range (40--50 dB), continuous frequency tuning and, very importantly, the possibility of \textit{direct} determination of spectra of any ``optical'' parameter, like complex conductivity, complex dielectric permittivity, etc. (`direct' means that no Kramers--Kronig analysis---typical for far-infrared Fourier transform spectroscopy---is needed). The principle of operation of BWO-spectrometers is described in details in the literature (see, e.g., \cite{6-Kozlov-Volkov, 7-Gorshunov-BWO_spectroscpoy}). It is based on measurement of the complex transmission coefficient $Tr^{*} = Tr\exp(i\varphi)$ of a plane-parallel sample with subsequent calculation of the spectra of its optical parameters from those of the transmission coefficient amplitude $Tr(\nu)$ and the phase $\varphi(\nu)$. The corresponding expression can be written as \cite{8-Born-Wolf, 9-Dressel}
\begin{equation}
Tr^{*}=Tr\exp(i\varphi) = \frac{T_{12}T_{21}\exp(i\delta)}{1+ T_{12}T_{21}\exp(2i\delta)}.\label{eqn:THz-Eq1}
\end{equation}
Here 
\begin{eqnarray}
\nonumber T_{pq}= t_{pq}\exp(i\varphi_{pq}), t^{2}_{pq}=\frac{4(n_{p}^2+k_{p}^2)}{(k_p + k_q)^2+(n_p + n_q)^2}, \varphi_{pq}= \arctan\{\frac{k_pn_p-k_qn_q}{n_p^2+k_p^2+n_pn_q+k_pk_q}\}
\end{eqnarray} 
are Fresnel coefficients for the interfaces `air--sample', indices $p,~ q = 1,~ 2 $ correspond: `1' to air (refractive index $n_1 = 1$, extinction coefficient $k_1 = 0$) and `2' to the material of the sample $(n_2,~k_2)$, $\delta = \frac{2{\pi}d}{\lambda}(n_2+ik_2)$, $d$ is the sample thickness, $\lambda$ is the radiation wavelength. The sample parameters (for instance, $n_2$ and $k_2$ ) are found for each fixed frequency by solving two coupled equations for the two unknowns, $Tr(n_2, k_2, \nu) = Tr_{\mathrm{exp}}(\nu)$ and $\varphi(n_2, k_2, \nu) =  \varphi_{\mathrm{exp}}(\nu)$ [here $Tr_{\mathrm{exp}}(\nu)$ and $\varphi_{\mathrm{exp}}(\nu)$ are the measured quantities]. The so-found values of $n_2(\nu)$ and $k_2(\nu)$ can then be used to derive the spectra of the complex permittivity $\varepsilon^*(\nu) = \varepsilon'(\nu) + i \varepsilon''(\nu) = n_2^2 - k^2_2 + 2 i n_2 k_2$, complex conductivity $\sigma^*(\nu) = \sigma_1(\nu) + i \sigma_2(\nu) = \nu n_2 k_2 + i \nu (\varepsilon_{\infty} - \varepsilon')/2$, etc. ($\varepsilon_{\infty}$ is the high-frequency contribution to the permittivity).
\pb

If the sample is characterized by low enough absorption coefficient, Fabry--Perot-like interference of the radiation within the plane-parallel layer leads to an interference maxima and minima in the transmission coefficient spectra. In this case there is no need to measure the phase shift spectra since the pairs of optical quantities of the sample can be calculated from the transmission coefficient spectrum alone: the absorptive part (like $\varepsilon''$ or $\sigma_1$) is determined from the interferometric maxima amplitudes and the refractive part (like $\varepsilon'$ or $n$) is calculated from their positions \cite{6-Kozlov-Volkov, 7-Gorshunov-BWO_spectroscpoy}.
\pb

When measuring the dielectric response of the films (like heterostructures in the present case) on dielectric substrates, first the dielectric properties of the substrate material are determined by standard techniques just described. Next, one measures the spectra of the transmission coefficient and of the phase shift of the film-substrate system, and it is these spectra that are used to derive the dielectric response of the film by solving two coupled equations for two unknowns---``optical'' parameters of the film. The corresponding expression for the complex transmission coefficient of a two-layer system can be written as  \cite{8-Born-Wolf, 9-Dressel}:
\begin{equation}
Tr^*_{1234}= Tr\exp(i\varphi)= \frac{T_{12}T_{23}T_{34}\exp\{i (\delta_2+\delta_3)\}}{1+T_{23}T_{34}\exp(2i\delta_3)+T_{12}T_{23}\exp(2i\delta_2)+ T_{12}T_{34}\exp\{2i(\delta_2+\delta_3)\}}, \label{eqn:THz-Eq2}
\end{equation}
where indices 1 and 4 refer to the media on the two sides of the sample, i.e., of the film on substrate, $\delta_p = (n_p + i k_p)$, with $d_p$ being the film and substrate thicknesses ($p = 2,~ 3$). The other notations are the same as in Eq.~(\ref{eqn:THz-Eq1}).
The measurements are performed in a quasioptical configuration, no waveguides are used \cite{6-Kozlov-Volkov, 7-Gorshunov-BWO_spectroscpoy} and this makes measurement schemes extremely flexible. All measurement and analysis procedures are PC-controlled. Most important parameters of the BWO-spectrometer are summarized in Table~\ref{tab:BWO_parameters}.

\section*{Results and Discussion}

\subsection*{Morphology and defects}

\subsubsection*{STM and RHEED study of Ge/Si(001) QD arrays: morphology and  formation}

Previously, we have shown in a number of {\textit{STM studies}} \cite{classification, VCIAN2011, Nucleation_high-temperatures, Hut_nucleation, initial_phase, CMOS-compatible-EMRS} that the process 
of the hut array nucleation and growth at low temperatures starts  from occurrence of two types of  16-dimer nuclei \cite{Hut_nucleation} on wetting layer (WL) patches of 4-ML height \cite{initial_phase} giving rise to two known species of $\{105\}$-faceted clusters---pyramids and wedges \cite{classification}---which then, growing in height (both types) and in length (wedges), gradually occupy the whole wetting layer, coalless and start to form a nanocrystalline Ge film (Figure~\ref{fig:STM-360}) \cite{VCIAN2011,CMOS-compatible-EMRS}. This is a life cycle of hut arrays at the temperatures \textless\,600\textcelsius. We refer to cluster growth at these temperatures as the low-temperature mode.
\pb

At high temperatures (\textgreater\,600\textcelsius), only pyramids represent a family of huts: they were found to nucleate on the WL patches in the same process of 16-dimer structure occurrence as at low temperatures \cite{Nucleation_high-temperatures}. We failed to find wedges or their nuclei if Ge was deposited at these temperatures and  this fact waits for a  theoretical explanation.
\pb

In addition to pyramids, shapeless Ge heaps faceting during annealing have been observed on WL in the vicinity of pits and interpreted as possible precursors of large faceted clusters \cite{VCIAN2011, Nucleation_high-temperatures}.
Note that a mechanism of Ge hut formation via faceting of some shapeless structures appearing near WL irregularities, which resembles the process described in the current article, was previously considered as the only way of Ge cluster nucleation on Si(001) \cite{Nucleation,Goldfarb_2005}. Now we realize that huts nucleate in a different way\cite{Hut_nucleation} and formation of the faceting heaps at high temperatures is a process competing with appearance of real pyramidal huts which arise due to formation of the 16-dimer nuclei on tops of WL patches \cite{Hut_nucleation,CMOS-compatible-EMRS,initial_phase}. Yet, further evolution of the Ge heaps into finalized faceted clusters, such as domes, in course of Ge deposition is not excluded \cite{Nucleation_high-temperatures}. 
\pb

During further growth at high temperatures, pyramids reach large sizes becoming much greater than their low-temperature counterparts and usually form incomplete facets or split edges (Figure~\ref{fig:STM-600}). An incomplete facet seen in Figure~\ref{fig:STM-600}a and especially a ``pelerine'' of multiple incomplete facets seen in  Figure~\ref{fig:STM-600}b,c around the pyramid top indicate unambiguously that this kind of clusters grow from tops to bottoms completing facets rather uniformly from apexes to bases, and bottom corners of facets are filled  the latest. Sometimes it results in   edge splitting  near the pyramid base (Figure~\ref{fig:STM-600}b,d).
\pb

{\it RHEED} has allowed as to carry our {\textit{in-situ}} explorations of forming cluster arrays. We have compared RHEED patterns of Ge/Si(001) surfaces during Ge deposition at different temperatures and a dynamics of diffraction patterns during sample heating and cooling.
\pb

 Diffraction patterns of reflected high-energy electrons for samples of thin ($h_{\rm Ge}=$ 4\,\AA) Ge/Si(001) films deposited at high (650 or 600\textcelsius) and low (360\textcelsius) temperatures with equal effective thicknesses   are presented in Figure~\ref{fig:rheed}a,b. The patterns are similar and represent a typical $(2\times 1)$ structure of Ge WL; reflexes associated with appearance of huts (the 3D-reflexes) are absent in both images, that agrees with the data of the STM analysis. Diffraction patterns presented in Figure~\ref{fig:rheed}a,c,e are related to the samples with $h_{\rm Ge}$ increasing from 4 to 6\,\r{A}. The 3D-reflexes are observed only in the pattern of the samples with $h_{\rm Ge}=$ 6\,{\AA}, that is also in good agreement with the STM data  \cite{VCIAN2011,initial_phase}.
\pb

Influence of the sample annealing at the  deposition temperature is illustrated by a complimentary pair of the RHEED patterns given in Figure~\ref{fig:rheed}c,d. Annealing of specimens at the temperature of growth (650\textcelsius) resulted in appearance of the 3D-reflexes (Figure~\ref{fig:rheed}d) that also corresponds with the results of our STM studies \cite{VCIAN2011}.
\pb

Difference in evolution of diffraction patterns during the deposition of Ge is a characteristic feature of the high-temperature mode of growth in comparison with the low-temperature one. The initial Si(001) surface before Ge deposition is $(2\times 1)$ reconstructed. At high temperatures, as $h_{\rm Ge}$ increases, diffraction patterns evolve as $(2\times 1)\rightarrow$ $(1\times 1)\rightarrow$ $(2\times 1)$ with very weak {\textonehalf}-reflexes. Brightness of the {\textonehalf}-reflexes gradually increases (the $(2\times 1)$ structure becomes pronounced) and the 3D-reflexes arise only during sample cooling (Figure~\ref{fig:RHEED_cool-600}). At low temperatures, the RHEED patterns change as $(2\times 1)\rightarrow$ $(1\times 1)\rightarrow$ $(2\times 1)\rightarrow$ $(2\times 1)+3$D-re\-f\-le\-xes. The resultant pattern does not change during sample cooling. 
\pb

This observation reflects the process of Ge cluster ``condensation'' from the 2D gas of mobile Ge adatoms. High Ge mobility and low cluster nucleation rate in comparison with fluxes to competitive  sinks of adatoms determines the observed difference in the surface structure formation at high temperatures as compared with that at low temperatures \cite{VCIAN2011,Nucleation_high-temperatures} when the adatom flux to nucleating and growing clusters predominates and adatom (addimer) mobility is relatively small.
\pb

\subsubsection*{STM and HRTEM study of Ge/Si heterostructures with QD array: morphology and defects}

Structures overgrown with Si were examined by means of HRTEM for structural perfection or possible defects, e.g., imperfections induced by array defects reported in Ref.~\cite{defects_ICDS-25}. 
\pb

Data of HRTEM studies evidence that  extended defects do not arise at low $h_{\rm Ge}$  on the buried Ge clusters
and perfect epitaxial heterostructures with quantum dots form under these conditions that enables the formation of  defectless  multilayer structures suitable for device applications. Figure~\ref{fig:TEM-6A} relates to the five-layer Ge/Si structure with {\textit{h$_{\mathrm{Ge}}$}}\,= 6\,\r{A}. We succeeded to resolve separate Ge clusters whose height is, according to our STM data \cite{VCIAN2011,classification}, $\apprle$\,3\,ML over  WL patches (Figure~\ref{fig:TEM-6A}a,b). A lattice structure next to the cluster apex is not disturbed (Figure~\ref{fig:TEM-6A}c,d); its parameters estimated from the Fourier transform of an image taken from this domain (Figure~\ref{fig:TEM-6A}e,f), $\sim 5.4$\,\r{A} along the [001] direction and $\sim 3.8$\,\r{A} along [110], within the accuracy of measurements coincide with the parameters of the undisturbed Si lattice.
\pb

Stacking faults (SF) have been found to arise above Ge clusters at $h_{\rm Ge}$ as large as 10\,\r{A} (Figure~\ref{fig:TEM-10A1L}). 
SFs often damage Si  structures with overgrown Ge layers at this   values of $h_{\rm Ge}$.  A high  perfection structure is observed around Ge clusters in Figure~\ref{fig:TEM-10A1L}a although their height is up to 1.5 nm over WL (the typical height of huts is known from  both our STM  and HRTEM data). Yet, a tensile strained domain containing such extended defects as SFs and twin boundaries  forms over a cluster shown in Figure~\ref{fig:TEM-10A1L}b,c (twinning is clearly observable in Figure~\ref{fig:TEM-10A1L}d). One can see, however, that this cluster is extraordinary high: its height over WL exceeds 3.5 nm. Such huge clusters have  been described by us previously as defects of arrays \cite{defects_ICDS-25}; we predicted in that article such formations to be able to destroy Ge/Si structures generating high stress fields in  Si spacer layers and, as a consequence, introducing extended defects in device structures. As seen in Figure~\ref{fig:TEM-10A1L}b,c, the stress field spreads under the cluster in the Si buffer layer grown at much higher temperature than the cap.
Unfortunately, the huge Ge hut clusters (as we showed in Ref.~\cite{defects_ICDS-25}, they are not domes) usually appear in the arrays and their number density was estimated as $\sim$\,$10^9$\,cm$^{-3}$ from the STM data.
\pb

Strain domains are also seen next to Ge clusters in the five-layer structures depicted in Figure~\ref{fig:TEM-9-10A} ({\textit{h$_{\mathrm{Ge}}$}}\,= 9 or 10\,\r{A}). We found that such domains are not inherent to all cluster vicinities but only to some of them (Figure~\ref{fig:TEM-9-10A}a,d). The disturbed strained domains give a contrast different from that of the undisturbed Si lattice (Figure~\ref{fig:TEM-9-10A}e). Zoom-in micrographs of the disturbed regions show perfect order of atoms in the crystalline lattice (Figure~\ref{fig:TEM-9-10A}b,c,e,f) everywhere except for the closest vicinities of the Si/Ge interface where point defects and a visible lattice disordering immediately next to the cluster are registered (Figure~\ref{fig:TEM-9-10A}b,c,e,f). However, some farther from the interfaces but still near cluster apexes the crystalline order restores (Figure~\ref{fig:TEM-9-10A}h). We have estimated the lattice parameter in the disturbed regions from the Fourier transforms of the HRTEM micrographs taken in these domains (Figure~\ref{fig:TEM-9-10A}i). The values we obtained appeared to be vary for different regions. Yet, they usually appreciably exceeded the Si lattice parameter. Moreover, they often reached the Ge parameter of $\sim$\,5.6--5.7\,\r{A} (along [001] and $\sim$\,4\,\r{A} along [110]). This might be explained either by appreciable diffusion of Ge from clusters (previously, we have already reported an appreciable diffusion of Si in Ge clusters in analogous structures from covering Si layers grown at 530{\textcelsius} \cite{Raman_conf,our_Raman_en}) or by Si lattice stretching under the stress. Likely both factors acts. 
\pb

It is worth while emphasising  that the stretched domains usually do not contain extended defects, as it is seen from the HRTEM micrographs, except for the cases of array defects (huge clusters) like that demonstrated in Figure~\ref{fig:TEM-10A1L}. We suppose that the extended defects in these regions arise because the strain exceeds an elastic limit near huge clusters.  
\pb

Finally, we have tried to find out if huge clusters exists  in arrays of  {\textit{h$_{\mathrm{Ge}}$}}\,= 9\,\r{A} (Figure~\ref{fig:STM-9A}). We have been convinced that  even in rather uniform arrays large clusters (Figure~\ref{fig:STM-9A}e), which might generate considerable stress, are abundant and even huge ones (Figure~\ref{fig:STM-9A}d), which should produce lattice disordering (extended defects), are available.
Effect of such defects as huge clusters on device performance and a cause of their appearance in hut arrays  await further detailed studies.
\pb

\subsection*{Photo-emf of Ge/Si {\textit{p--i--n}}-structures}

\subsubsection*{Photo-emf spectra}

We have investigated heteroepitaxial Si {\textit{p--i--n}}-diodes with multilayer stacks of Ge/Si(001) QD dense arrays built in intrinsic domains  and found them to exhibit the photo-emf in a wide spectral range from 0.8 to 5 $\mu$m \cite{NES-2011,photon-2011}. An effect of wide-band  irradiation by infrared light on the photo-emf spectra has been observed. Here we describe the most representative data obtained for two radically different structures denoted as  R\,163 and R\,166 (Figure~\ref{fig:p-i-n_Schematics}).
\pb

Typical photo-emf spectra obtained for R\,163 and R\,166 structures are presented in Figure~\ref{fig:r163_r166}. In the spectra, we mark out three characteristic ranges which differently respond to bias lighting  and differently depend on its power.
\pb
      (i) {\textit{Wavelength range from 0.8 to 1.0 $\mu$m}.} The photo-emf response increases with the increase in the bias lighting power, reaches maximum at $P \approx 0.63$ mW/cm$^2$ with a Si filter and at $P \approx 2.6$ mW/cm$^2$ with a Ge filter and decreases with further increase in the power.\pb

      (ii) {\textit{Wavelength range from 1.1 to 2.6 $\mu$m}.} The photo-emf response decreases monotonously in this range with the   increase in the power of the bias lighting   with  any, Si or Ge, filter.\pb

     (iii) {\textit{Wavelength range \textgreater\,2.6 $\mu$m}.} The photo-emf response increases with the increase in the bias lighting power  and comes through its maximum at $P \approx 0.63$  mW/cm$^2$ if a Si filter is used and at $P \approx 0.25$ mW/cm$^2$ for a Ge filter. The response decreases with further growth of the bias lighting power  for a Si filter and remains unchanged when  Ge filter is utilized.\pb

We propose the following model for explanation of these observations:
In the studied structures, all QD layers are located in the $i$-domain 
(Figure~\ref{fig:bands}). One can see from these sketches that some QD layers are positively charged (the ground states of QDs is above the Fermi level and hence they are filled by holes) while others are neutral (the QDs' ground states are  below Fermi level and hence empty).
Then, one may consider a QD layer as a single ensemble of interacting centers because the average distance between QDs' apexes is about 13 nm whereas QDs' bases adjoin. Consequently, one can imagine an allowed energy band with some bandwidth, determined by QDs' sizes and composition dispersion, and a certain density of states in this band.
Let us explore in detail every  range of the photo-emf spectra  taking into account the proposed model.
\pb
      
\subsubsection*{Wavelength range from 0.8 to 1.0\,$\mu$m }

Without bias lighting, all radiation in the Si fundamental absorption range can be believed to be absorbed in  Si (cap-layer, spacers, buffer layer and substrate) and QDs are not involved in the absorption, so the total charge of QD layers remains unaltered. Electron-holes pairs are generated in the intrinsic region of the {\textit{p--i--n}}-diode as a result of the absorption and separated by the junction field which converts the radiation to emf. However, carrier separation is hindered because of presence of the potential barriers for holes in the valence band which are produced by the charged QD layers situated in intrinsic domain. Calculated height of these barriers equals 0.1 to 0.2 eV depending on the layer position in the structure.
\pb

Transitions from QD ensemble states to the valence and conduction bands of Si start under bias lighting. Carriers excited by bias lighting do not contribute to the photo-emf signal measured at the modulation frequency of the narrow-band radiation. QDs captured a photon change their charge state. An effective layer charge decreases as a result of the absorption of the bias lighting radiation  that results in reduction of potential barrier height  and more efficient carrier separation in the junction field. Increase in the photo-emf response  in the fundamental absorption range under bias lighting is explained by this process.

\subsubsection*{Wavelength range from 1.1 to 2.6\,$\mu$m }

This band is entirely below the Si fundamental absorption range. Therefore the response in this region cannot be explained in terms of absorption in bulk Si. One can explain the presence of the photo-emf signal in this region considering the following model: Both hole transitions from the QD ensemble states to the valence band and electron transitions from the QD ensemble states to the conduction band due to absorption of photons with the energy between $\sim 1.12$ and  $\sim 0.4$ eV are possible. The probability of every kind of the transitions is determined by the photon energy, the density of states in the QD ensemble  and  by effective charge of the QD layer. 
\pb

It follows from theoretical studies \cite{Gerasimenko_Si-mat_nanoelectr,Brudnyi-Ge-small_QD} 
and experiments on photoluminescence  \cite{PL-Si/Ge_1.4-1.8mcm,PL-Si/Ge} 
 that photons with energies ranged from 0.7 to 0.9 eV are required for electron transitions from the QD states to the conduction band. However, it is necessary to mention the research of photoconductivity \cite{Talochkin-Lateral_photoconductivity_Ge/Si}, 
in which electron transitions for low photon energy ($\sim 0.4$ eV) have been shown to be likely. The availability of these transitions is explained by dispersion of  sizes and composition of QDs, effect of diffusion on the hetero-interface  and deformation effects.
\pb

The likelihood of electron transitions drops rapidly with photons energy decrease because of reduction of the  density of states in the QDs ensemble  when approaching to the conduction band edge. This is the reason of the observed monotonous decrease in the photo-emf signal  with the increase in the radiation wavelength  in this range.
\pb

At the same time, bias lighting switching on leads to growth of concentration of the unmodulated (``dark'') carrier, depletion of QDs   and as a consequence to the observed reduction of the photo-emf response at the chopping frequency.
\pb

\subsubsection*{Wavelength range \textgreater\,2.6\,$\mu$m}

    As mentioned above, electron transitions can happen at low energy of the exciting radiation  ($\sim$\,0.4 eV) which correspond to wavelength of $\sim$\,3.1 $\mu$m. Yet, the photo-emf signal is observed at the radiation wavelengths up to 5 $\mu$m in our measurements. The presence of the photo-emf response in this range can  only be explained if   the QD layer is considered as a single ensemble of mutually interacting centers. An effective positive charge in the QD layer forms a potential well for electrons in the conduction band. This leads to reduction of energy needed for electron transitions from the QD ensemble states to the conduction band. Partial emptying of the states makes electron transitions possible and, at the same time, does not lead to significant change in the potential wells depth. As a result, electron transitions can happen at the exciting radiation energies as low as 0.25 eV. Hole transitions also can happen at these energies via a large number of excited states in the QD ensemble. 
\pb

It may be concluded that the likelihood the electron transitions  decreases faster than that of the hole transitions as the exciting radiation energy decreases in the considered wavelength range. 
However, first it is necessary  to empty the levels by the electron transitions  to make  possible hole transitions. 
This could be achieved by using an additional radiation of the spectral domain where the probability of the electron transitions is high. So, bias lighting stimulates the hole transitions by exciting electrons that leads to emptying the levels. In this case the electron concentration is not modulated as distinct from the hole concentration which is modulated at the chopper frequency. 
 This explains the observed low magnitude of the photo-emf  in the wavelength range \textgreater\,2.6 $\mu$m and its increase under bias lighting.

\subsubsection*{Influence of buffer layer thickness  on photo-emf spectra}

 As seen from Figure~\ref{fig:bands}, the buffer layer thickness determines the QD layers position the in intrinsic domain and thus controls the relative position  of the  Fermi level and the mini-band of the QD array in the region where the QD layers are situated. 
The charge of the QD layer is determined by  the band occupation of the QD ensemble  which, in turn, is controlled by the Fermi level location. 
For this reason the effect of bias lighting on  photo-emf generated by the narrow-band radiation in the fundamental absorption range is much stronger for the R\,166 structure, which have a thick buffer layer, than for the R\,163 one. This is clearly seen in Figure~\ref{fig:bias}. The absolute  value of photo-emf in the R\,166 structure is lower than that in the R\,163 sample due to higher potential barriers for holes in the valence band. Yet, the photo-emf response increases with the growth of the bias lighting power much stronger in the R\,166 {\textit{p--i--n}}-diode than in the  R\,163 one.

\subsubsection*{Prospective photovoltaic IR detectors}

On the basis of our results on the photo-emf in the Si {\textit{p--i--n}}-structures with Ge quantum dots, we have recently proposed \cite{Yur1-patent-Ge} a new design of photovoltaic quantum dot infrared photodetectors 
which enables detection of variations of photo-emf produced by the narrow-band radiation in the Si fundamental absorption range (a reference beam) under the effect of the wide-band IR radiation inducing changes in the Coulomb potential of the quantum dot ensemble which, in turn, affects the efficiency of the photovoltaic conversion of the reference beam. The quantum dot array resembles a grid of a triode in these detectors which is controlled by the detected IR light. The reference narrow-band radiation generates a potential between anode and cathode of this optically driven quantum dot triode; a magnitude of this voltage depends on the charge of the QD grid (Figure~\ref{fig:bands}). Such detectors can be fabricated on the basis of any appropriate semiconductor structures with potential barriers, e.g., {\textit{p--i--n}}-structures,  $p$--$n$-junctions or Schottky barriers, and built-in arrays of nanostructures.
\pb

There are many ways to deliver the reference beam to the detector, e.g., by irradiating the sensor by laser or LED.
We propose, however,  surface plasmon polaritons delivered to the detector structures by the plasmonic waveguides \cite{Bozhevolnyi-waveguides,Zayats-waveguides} to be applied as the reference beams in the detector circuits. This approach makes such detectors, if based on Si, fully compatible with existing CMOS fabrication processes \cite{Zayats-Si_waveguides} that, in turn, opens a way to development of plasmonic IR detector arrays on the basis of the monolithic silicon technology.
\pb

\subsection*{THz conductivity of multilayer Ge/Si QD arrays}

The effective dynamic conductivity of Ge quantum dot layer was determined by measuring the transmission coefficient spectra of heterostructures grown on Si(001) substrates. Characteristics of the substrates were determined beforehand as demonstrated by Figures~\ref{fig:THz-fig1} and \ref{fig:THz-fig2}. In Figure~\ref{fig:THz-fig1}, the interferometric pattern in the transmission coefficient spectrum $Tr(\nu)$ of a plane-parallel Si substrate is clearly seen. Pronounced dispersion of $Tr(\nu)$ peaks and their temperature dependence allow to extract the parameters of the charge carriers (holes) by fitting the spectra with Eq.~(\ref{eqn:THz-Eq2}) and by modelling the sample properties with the Drude conductivity model where the complex AC conductivity is given by an expression \cite{9-Dressel,10-Sokolov}
\begin{equation}
\sigma^*(\nu) = \sigma_1(\nu) + i\sigma_2(\nu) = \frac{\sigma_0\gamma^2}{\gamma^2+\nu^2} + i\frac{\sigma_0\nu\gamma}{\gamma^2+\nu^2}.
\label{eqn:THz-Eq3}
\end{equation}
Here $\sigma_1$ is the real part and $\sigma_2=\nu(\varepsilon_{\infty} - \varepsilon')/2$  is the imaginary part of the conductivity, $\varepsilon_{\infty}$  is the high-frequency dielectric constant, $\sigma_0=\nu^2_{\mathrm{pl}}/2\gamma$  is the DC conductivity, $\nu^2_{\mathrm{pl}} = (ne^2 /{\pi}m^*)^{\frac{1}{2}}$ is the plasma frequency of the carriers condensate, $n$, $e$ and $m^*$ are, respectively, their concentration, charge and effective mass and $\gamma$ is their scattering rate.  Figure~\ref{fig:THz-fig2}  shows the temperature variation of the  plasma frequency and the scattering rate of charge carriers. Lowering of the plasma frequency is mainly connected with the carriers' freezing out and the $\gamma(T)$ behaviour is well described by a $T^{-\frac{3}{2}}$ dependence, as expected.
\pb

The values of effective dynamical conductivity and absorption coefficient $\alpha = 4{\pi}k/\lambda$  of the heterostructures with Ge quantum dots were determined basing on the measurements of terahertz transmission coefficient spectra of the Si substrate with the heterostructure on it as compared to the spectra of the same substrate with the heterostructure etched away; this allowed us to avoid influence of (even slight) differences in dielectric properties of substrates cut of a standard commercial silicon wafer. By comparing the so-measured transmissivity spectra we reliably detect, although small, changes in the amplitudes of interference maxima of a bare substrate caused by heterostructures. This is demonstrated by Figure~\ref{fig:THz-fig3}: at $T = 300$\,K we clearly and firmly register a 2\% lowering of the peak transmissivity introduced by the heterostructure. When cooling down, the difference decreases and we were not able to detect it below about 170\,K, see Figure~\ref{fig:THz-fig4}. Correspondingly, as is seen in Figure~\ref{fig:THz-fig4}, the AC conductivity of the heterostructure decreases while cooling, along with the conductivity of the Si substrate. The latter observation might be an indication of the fact that the charges are delivered into the quantum dots array from the substrate; the statement, however, needs further exploration.
\pb

Measuring the room temperature spectra, we have found that the AC conductivity and the absorption coefficient of the heterostructure do not depend on the effective thickness (measured by the quartz sensors during MBE) of the germanium layer ($h_{\mathrm{Ge}}$) for $h_{\mathrm{Ge}}$ ranging from 8 to 14\,\r{A}, see Figure~\ref{fig:THz-fig5}. For larger coverage, $h_{\mathrm{Ge}}>14$\,\AA, both quantities start to decrease.
\pb

One of the main findings of this work is that the AC conductivity and absorption coefficient of Ge/Si heterostructures have been discovered to be significantly higher than those of the structure with the same amount of germanium not organized in an array of quantum dots. Crucial role played by quantum dots is supported by a decrease of $\sigma_{\rm AC}$ and $\alpha$ observed for large germanium coverage ($h_{\mathrm{Ge}}>14$\,\AA), when structurization into quantum dots gets less pronounced and the thickness of Ge layer becomes more uniform. On the other hand, it is worth noting that no extra absorption of terahertz radiation was detected in the samples with low coverage, $h_{\mathrm{Ge}}=4.4$ and 6\,\AA. This can be explained either by the absence of quantum dots in that thin Ge layer or by their small sizes, by a large fraction of the free wetting layer or by relatively large distances between the clusters as compared to their sizes, i.e., by the absence or smallness of the effect of quantum dots on the dielectric properties of the heterostructure.
\pb

As seen from Figure~\ref{fig:THz-fig5}, the values $\sigma_{\rm AC}\approx 100\,\Omega^{-1}\mathrm{cm}^{-1}$ and $\alpha\approx 4000\,\mathrm{cm}^{-1}$ are considerably higher than the values measured for bulk germanium, $\sigma_{\rm AC}({\rm Ge})\approx 10^{-2}\,\Omega^{-1}\mathrm{cm}^{-1}$, by about four orders of magnitude, and $\alpha({\rm Ge})\approx 40\,\mathrm{cm}^{-1}$, by about two orders of magnitude. Assuming that the AC conductivity of heterostructure is connected with the response of (quasi) free carriers, one can express it with a standard formula  $\sigma=e\mu n =ne^2(2\pi \gamma m^*)^{-1}$ ($\mu$ is the mobility of charge carriers). Then, the observed increase has to be associated with considerable enhancement either of the mobility (suppression of scattering rate) of charge carriers within a quantum dot array or of their concentration. The second possibility has to be disregarded since the total concentration of charges in the sample (substrate plus heterostructure) remains unchanged. As far as the mobility increase is concerned, we are not aware of a mechanism that could lead to its orders of magnitude growth when charges get localized within the quantum dot array.
\pb

Another interpretation of the observed excess AC conductivity could be based on some kind of {\textit{resonance}} absorption of terahertz radiation. Known infrared experiments exhibit resonances in quantum dot arrays that are caused by the transitions between quantized energy levels, as well as between the split levels and the continuum of the valence or conduction band \cite{4-Drexler-InGaAs,11-Heitmann,12-Boucaud,13-Weber,14-Savage}. Carriers localized within quantum dots can form bound states with the carriers in the surrounding continuum (excitons) or with optical phonons (polarons), which can in turn interact with each other and form collective complexes \cite{3-Colomb_interactions-Dvur,4-Drexler-InGaAs,12-Boucaud,13-Weber,14-Savage,15-Hameau}. Plasma excitations generated by electromagnetic radiation in the assembly of conducting clusters or quantum dots also have energies of about 10 meV \cite{16-Sikorski,18-Dahl,17-Demel}, i.e., fall into the THz band. It is important that these effects can be observed not only at low, but at elevated temperatures as well, up to the room temperature. At this stage, we are not able to unambiguously identify the origin of the THz absorption seen at $T = 170$ to 300\,K in Ge/Si heterostructure with Ge quantum dots. Among the aforementioned, the mechanisms involving polaritons or plasma excitations seem to be least affected by thermal fluctuations and could be considered as possible candidates. To get detailed insight into microscopic nature of the observed effect, further investigations of heterostructures with various geometric and physical parameters, as well as in a wider frequency and temperature intervals are in progress.
\pb

\section*{Conclusions}

In conclusion of the article, we highlight its main provisions.
\pb
Using high resolution STM and {\it in-situ} RHEED we have explored the processes of Ge hut cluster array formation and growth at low temperatures on the Ge/Si(001) wetting layer. 
 Different dynamics of the RHEED patterns in the process of Ge hut array formation at low and high temperatures of Ge deposition reflects the difference in adatom mobility and their  fluxes from 2D gas of mobile particles (atoms, dimers and dimer groups) on the surface which  govern the nucleation rates and densities of arising Ge clusters.
\pb

HRTEM studies of multilayer Ge/Si heterostructures with buried arrays of Ge huts have shown that the domains of stretched lattice  occurring over Ge clusters in Si layers at high Ge coverages  usually do not contain extended defects. We suppose that the extended defects in these regions arise because the strain exceeds an elastic limit near huge clusters.  
\pb

Silicon {\textit{p--i--n}}-diodes with multilayer stacks of Ge cluster arrays built in {\textit{i}}-domains have been found to exhibit the photo-emf in a wide spectral range from 0.8 to 5\,$\mu$m.  A significant increase in photo-emf response  in the fundamental absorption range under the wide-band IR radiation has been reported and explained in terms of positive and neutral charge states of the quantum dot layers and the Coulomb potential of the quantum dot ensemble. A new type of photovoltaic QDIPs is proposed in which  photovoltage generated by a reference beam in the fundamental absorption band is controlled by the QD grid charge induced by the detected IR radiation \cite{Yur1-patent-Ge}.
\pb

Using a BWO-spectrometer, first measurements of terahertz dynamical conductivity spectra of Ge/Si heterostructures were carried out at frequencies ranged from 0.3 to 1.2 THz in the temperature interval from 5 to 300\,K. The effective dynamical conductivity of the heterostructures with Ge quantum dots has been found to be significantly higher than that of the structure with the same amount of Ge not organized in  quantum dots. The excess conductivity is not observed in the structures with the Ge coverage less than 8\,\AA. When a Ge/Si sample is cooled down the conductivity of the heterostructure decreases.
\pb

\section*{Abbreviations}
AC, alternating current;
BWO, backward-wave oscillator;
CMOS, complementary metal-oxide semiconductor; 
CZ, Czochralski or grown by the Czochralski method;
DC, direct current;
emf, electromotive force;
HRTEM, high resolution transmission electron microscope; 
IR; infrared;
LED, light emitting diode;
MBE, molecular beam epitaxy;      
ML, monolayer; 
QD, quantum dot;  
QDIP, quantum dot infrared photodetector;  
RHEED, reflected high energy electron diffraction; 
SF, stacking fault;
SIMS, secondary ion mass spectroscopy; 
STM, scanning tunneling microscope;  
WL, wetting layer; 
UHV, ultra-high vacuum.
\pb

\section*{Competing interests}
The authors declare that they have no competing interests.  
\pb
    
\section*{Authors contributions}
VAY conceived of the study and designed it, performed data analysis, and took part in discussions and interpretation of the results; he also supervised and coordinated the research projects. 
LVA participated in the design of the study, carried out the experiments, performed data analysis, and took part in discussions and interpretation of the results. 
MSS investigated the photo-emf spectra; he carried out the experiments, performed data analysis, and took part in discussions and interpretation of the results. 
VAC participated in the design of the study, took part in discussions and interpretation of the results; he also supervised the researches performed by young scientists and students. 
KVC took pat in the experiments on investigation of the photo-emf spectra and the terahertz conductivity; he prepared experimental samples and took part in discussions and interpretation of the results. 
OVU performed the HRTEM studies and took part in discussions and interpretation of the results. 
VPK participated in the design of the study, took part in discussions and interpretation of the results; he also supervised the research project. 
ESZ carried out the experiments on the terahertz spectroscopy; she performed measurements and data analysis, and took part in discussions and interpretation of the results. ASP participated in the studies by the terahertz spectroscopy; he took part in discussions and interpretation of the results. 
IES participated in the studies by the terahertz spectroscopy; he took part in discussions and interpretation of the results. 
BPG performed the explorations by the terahertz spectroscopy; he participated in the design of the study, performed measurements and data analysis, and took part in discussions and interpretation of the results; he also supervised the research project.
\pb

\section*{Acknowledgements}
  \ifthenelse{\boolean{publ}}{\small}{This research has been supported by the Ministry of Education and Science of RF through the contracts No.~14.740.11.0069 and 16.513.11.3046 and by RFBR through the grant No.~11-02-12023-ofi-m. Equipment of the Center of Collective Use of Scientific Equipment of GPI RAS was utilized for this study. We acknowledge the financial and technological support of this work.}
\pb





{\ifthenelse{\boolean{publ}}{\footnotesize}{\small}
 \bibliographystyle{bmc_article}  
  \bibliography{EMNC2012-article} }     


\ifthenelse{\boolean{publ}}{\end{multicols}}{}





\clearpage

\section*{Tables}
  \subsection*{Table~\ref{tab:BWO_parameters} - Main parameters of the terahertz BWO-spectrometer}
\begin{table}[h]
    \par
    \mbox{
      \begin{tabular}{|l|l|}
       \hline 
       Frequency range, THz & 0.03 to 2   \\ \hline
       Probing radiation power, mW & 1 to 100 \\ \hline
        \multicolumn{2}{|l|}{Frequency resolution: }  \\ \hline
~~~~~relative,  $\Delta\nu /\nu$&    $10^{-4}$--$10^{-5}$        \\ \hline
~~~~~absolute, cm$^{-1}$ &  0.001   \\ \hline
Dynamical range, dB &   40 to 60   \\ \hline
Signal to noise ratio &      $10^4$ to $10^6$      \\ \hline
Probing radiation polarization degree, \% &   99.99        \\ \hline
 \multicolumn{2}{|l|}{Time to record a spectrum of 100 points, s:}   \\ \hline
~~~~~amplitude $Tr(\nu)$   &  10 to 20  \\ \hline
~~~~~phase $\varphi(\nu)$    &    20 to 40   \\ \hline
Temperature interval, K      &   2 to 300          \\ \hline
Magnetic fields, T         &         up to 7       \\ \hline
      \end{tabular}
      }

\caption{\label{tab:BWO_parameters}
}
\end{table}

\clearpage

\section*{Figures}

\begin{figure*}[h]
\includegraphics[scale=1.2]{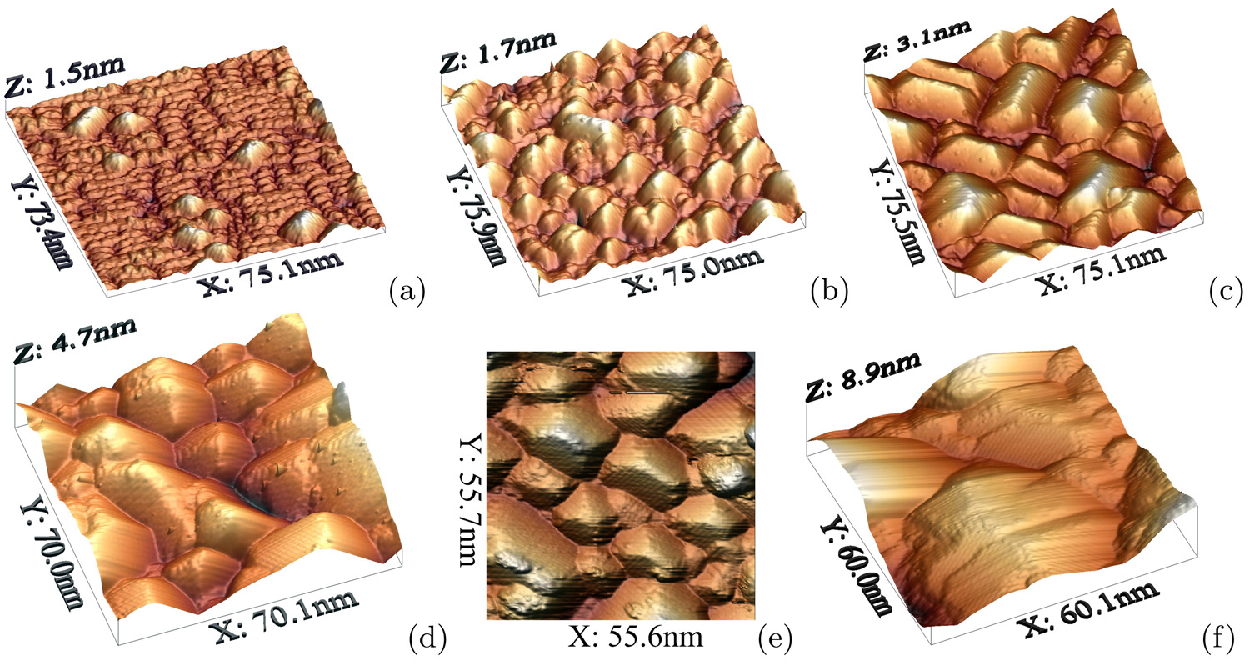}
\caption{\label{fig:STM-360}
}
\end{figure*}
\subsection*{Figure~\ref{fig:STM-360} - 
STM images of Ge/Si(001) quantum dot arrays grown at 360\textcelsius:
}
$h_{\rm Ge}$ (\AA) is
(a) 6, 
(b) 8, 
(c) 10, 
(d) 14, 
(e) 15, 
(f) 18.
\pb  

\clearpage
\begin{figure*}
\includegraphics[scale=1]{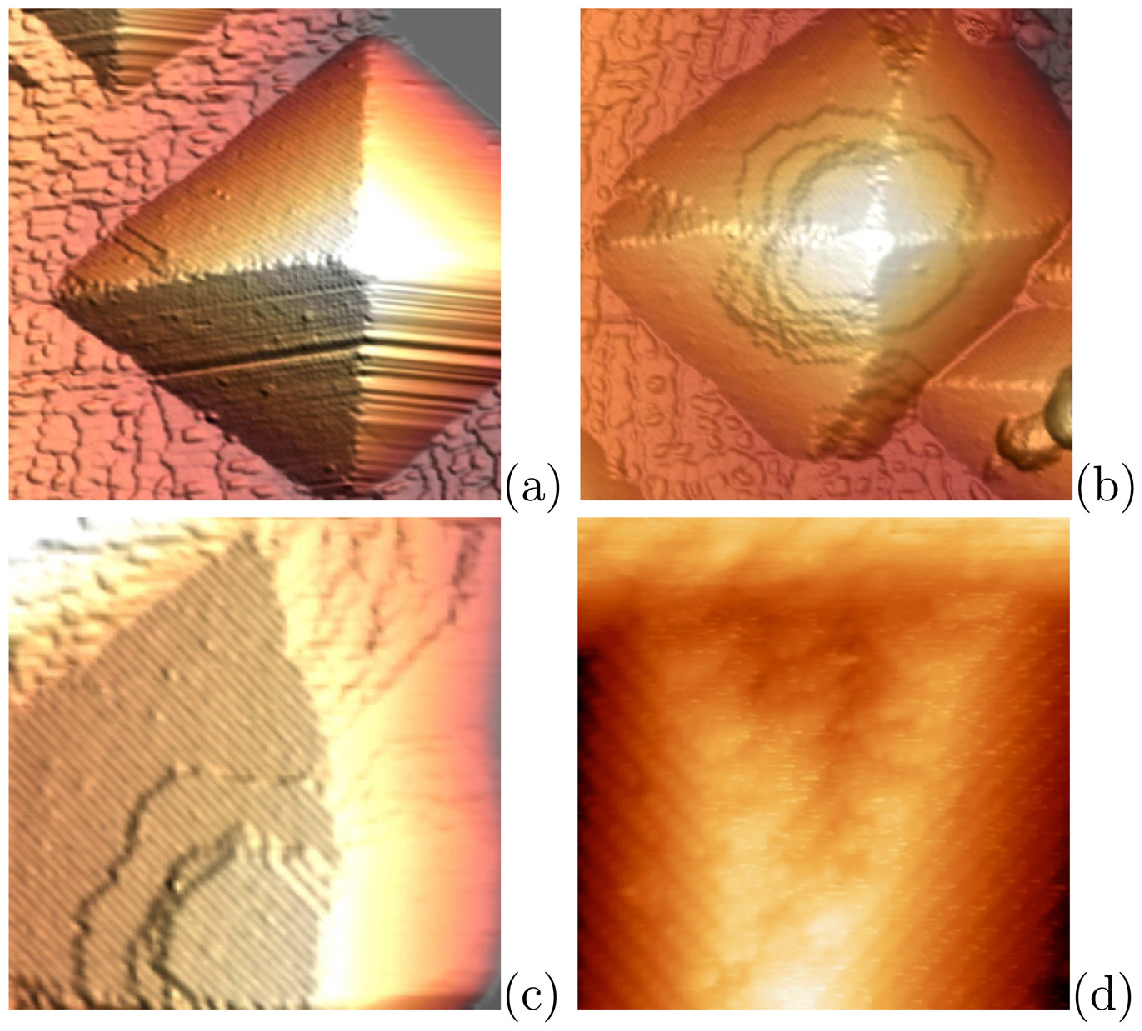}
\caption{\label{fig:STM-600}
}
\end{figure*}
\subsection*{Figure~\ref{fig:STM-600} - 
STM empty-state images of high-temperature pyramids:
}
$T_{\rm gr}=650$\textcelsius;
(a) $87\times87$\,nm, steps of the incomplete upper left facet, running normal to the base side, are seen near the left corner of the pyramid;
(b) $87\times87$\,nm, a cluster with edges split near the base and an apex formed by a set of incomplete \{105\} facets;
(c) $57\times57$\,nm, a magnified image of a facet with several \{105\} incomplete facets near an apex;
(d) $22\times22$\,nm, a split edge near a  base.
\pb

\clearpage
\begin{figure*}
\includegraphics[scale=1.2]{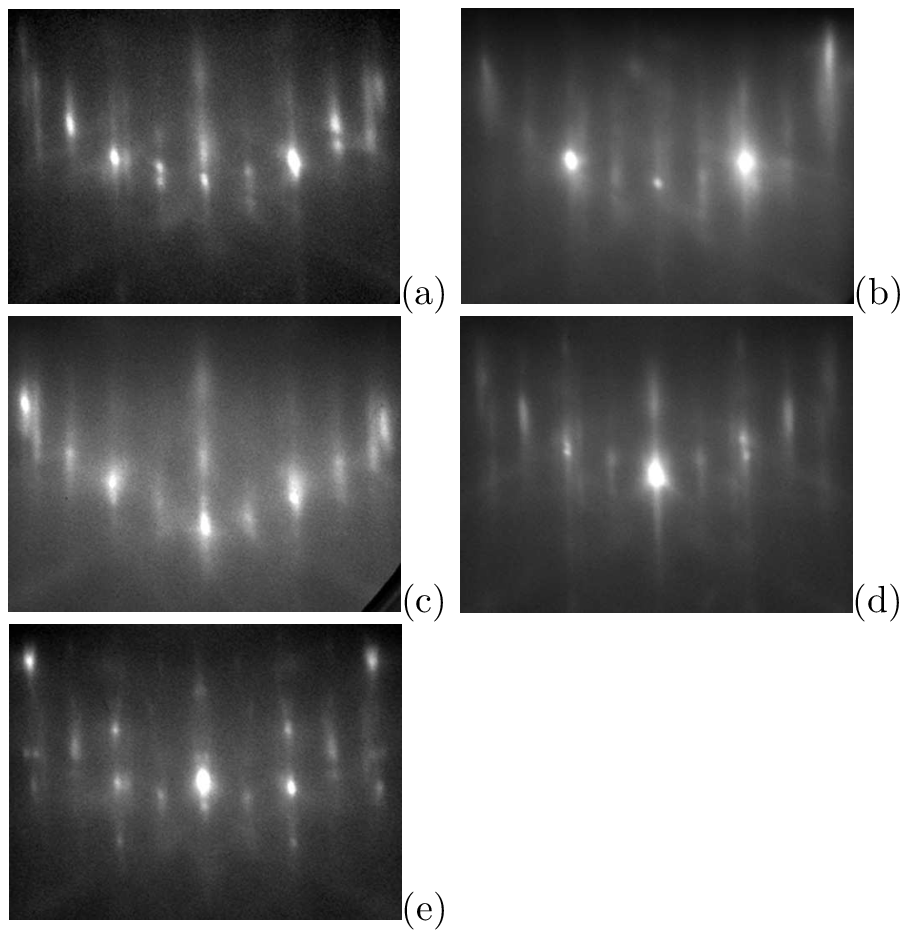}
\caption{\label{fig:rheed}
}
\end{figure*}
\subsection*{Figure~\ref{fig:rheed} - 
\textit{In situ}
RHEED patterns of Ge/Si(001) films:
}
\textit{E} = 10\,keV, [110] azimuth;
(a) $T_{\rm gr} =$ 650{\textcelsius}, $h_{\rm Ge}=$ 4\,\AA;
(b) $T_{\rm gr} =$ 360{\textcelsius}, $h_{\rm Ge}=$ 4\,\AA;
(c) $T_{\rm gr} =$ 650{\textcelsius}, $h_{\rm Ge}=$ 5\,\AA;
(d) $T_{\rm gr} =$ 650{\textcelsius}, $h_{\rm Ge}=$ 5\,\AA, annealing  at the deposition temperature for 7 min;
(e) $T_{\rm gr} =$ 650{\textcelsius}, $h_{\rm Ge}=$ 6\,\AA, the similar pattern is obtained for $T_{\rm gr} =$ 600{\textcelsius}; 
the patterns were obtained at room temperature after sample cooling.
\pb

\clearpage
\begin{figure*}
\includegraphics[scale=.7]{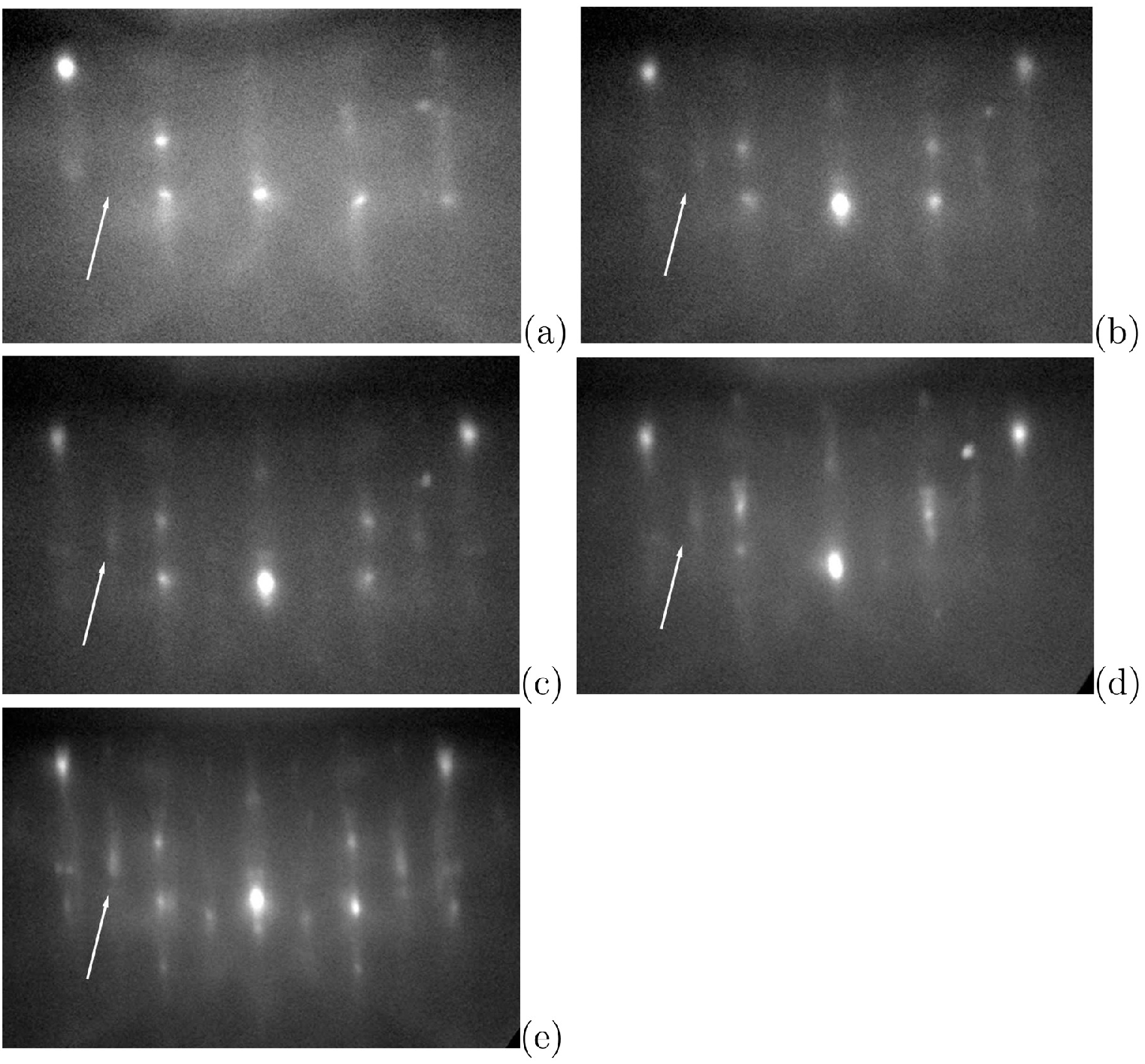}
\caption{\label{fig:RHEED_cool-600}
}
\end{figure*}
\subsection*{Figure~\ref{fig:RHEED_cool-600} - 
RHEED patterns of Ge/Si(001) deposited at 600{\textcelsius} obtained during sample cooling:
}
$h_{\rm Ge}=$ 6\,\AA;
\textit{E} = 10\,keV, [110] azimuth;
cooling rate is $\sim 0.4$\textcelsius/s (see the cooling diagram in Ref.~\cite{stm-rheed-EMRS});
(a)  $T=600$\textcelsius, before cooling;
(b)--(d) during cooling, time from beginning of cooling (min.):
(b) 1,
(c) 2,
(d) 3;
(e) room temperature, after cooling;
arrows indicate  the arising {\textonehalf}-reflexes to demonstrate a process of the $(2\times 1)$ pattern appearance; 
the images were cut from frames of a film.
\pb

\clearpage
\begin{figure*}[h]
\includegraphics[scale=.8]{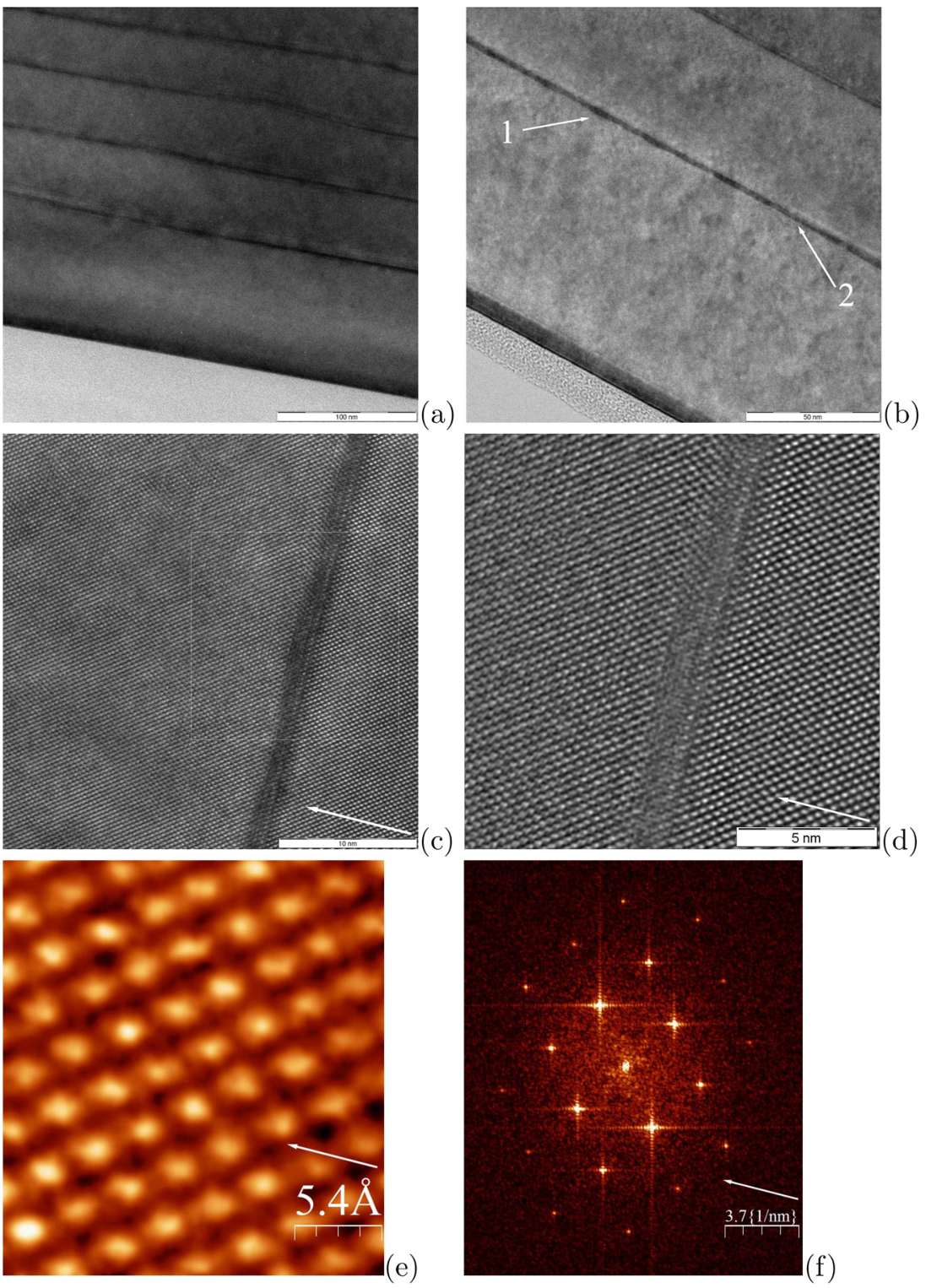}
\caption{\label{fig:TEM-6A}
 }
\end{figure*}
\subsection*{Figure~\ref{fig:TEM-6A} - 
HRTEM data for the five-layer Ge/Si  heterostructure with buried Ge clusters:
}
{\textit{h$_{\mathrm{Ge}}$}}\,= 6\,\r{A} (see Figure~\ref{fig:STM-360}a);
(a) a long shot, the mark is 100 nm;
(b) Ge clusters resolved in a layer, figure `1' indicates one of the clusters, `2' shows a WL segment; the mark is 50 nm;
(c),(d) magnified images of a Ge cluster, the panel (d) corresponds to the light square in the panel (c); the marks are 10 and 5 nm, respectively;
(e)  a close-up image of a domain next to the top of the cluster imaged in (d);
(f) the Fourier transform of the image (e), the measured periods are $\sim 5.4$\,\r{A} along [001] and $\sim 3.8$\,\r{A} along [110];
arrows in panels (c) to (f) indicate the [001] direction.
\pb

\clearpage
\begin{figure*}[h]
\includegraphics[scale=1]{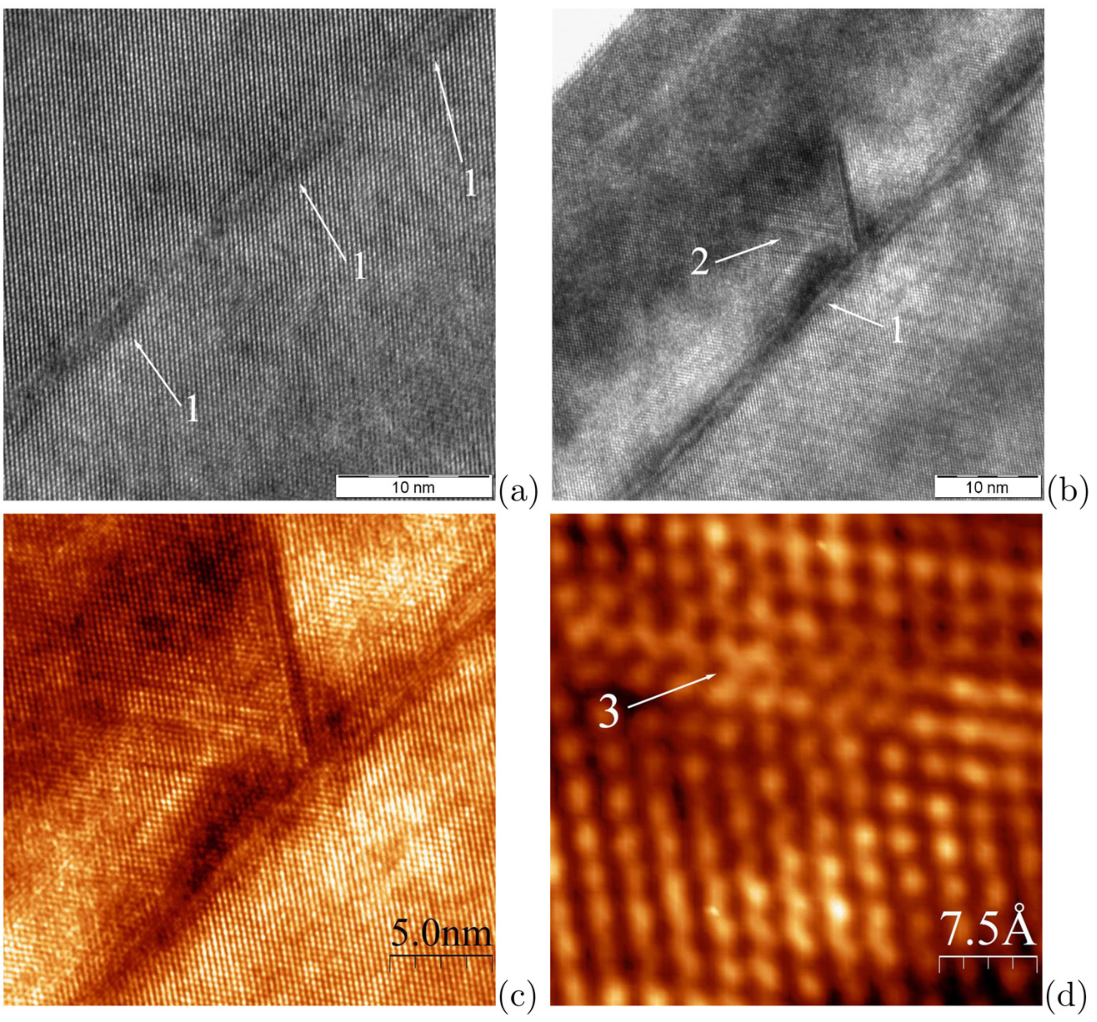}
\caption{\label{fig:TEM-10A1L}
}
\end{figure*}
\subsection*{Figure~\ref{fig:TEM-10A1L} - 
HRTEM images of the one-layer Ge/Si structures with buried Ge clusters:
}
{\textit{h$_{\mathrm{Ge}}$}}\,= 10\,\r{A} (see Figure~\ref{fig:STM-360}c);
(a) a perfect epitaxial structure of Ge  and Si layers; the mark is 10 nm;
(b),
(c) a huge cluster (\textgreater\,3,5 nm high) gives rise to tensile strain generating point and extended defects in the Si cap, the stress field spreads under the cluster [the mark is 10 nm in (b)];
(d) a magnified image obtained from the tensile domain, extended defects are seen;
`1' denotes Ge clusters, `2' is a domain under tensile stress, `3' indicates a twin boundary.
\pb

\clearpage
\begin{figure*}[h]
\large
\includegraphics[scale=.70]{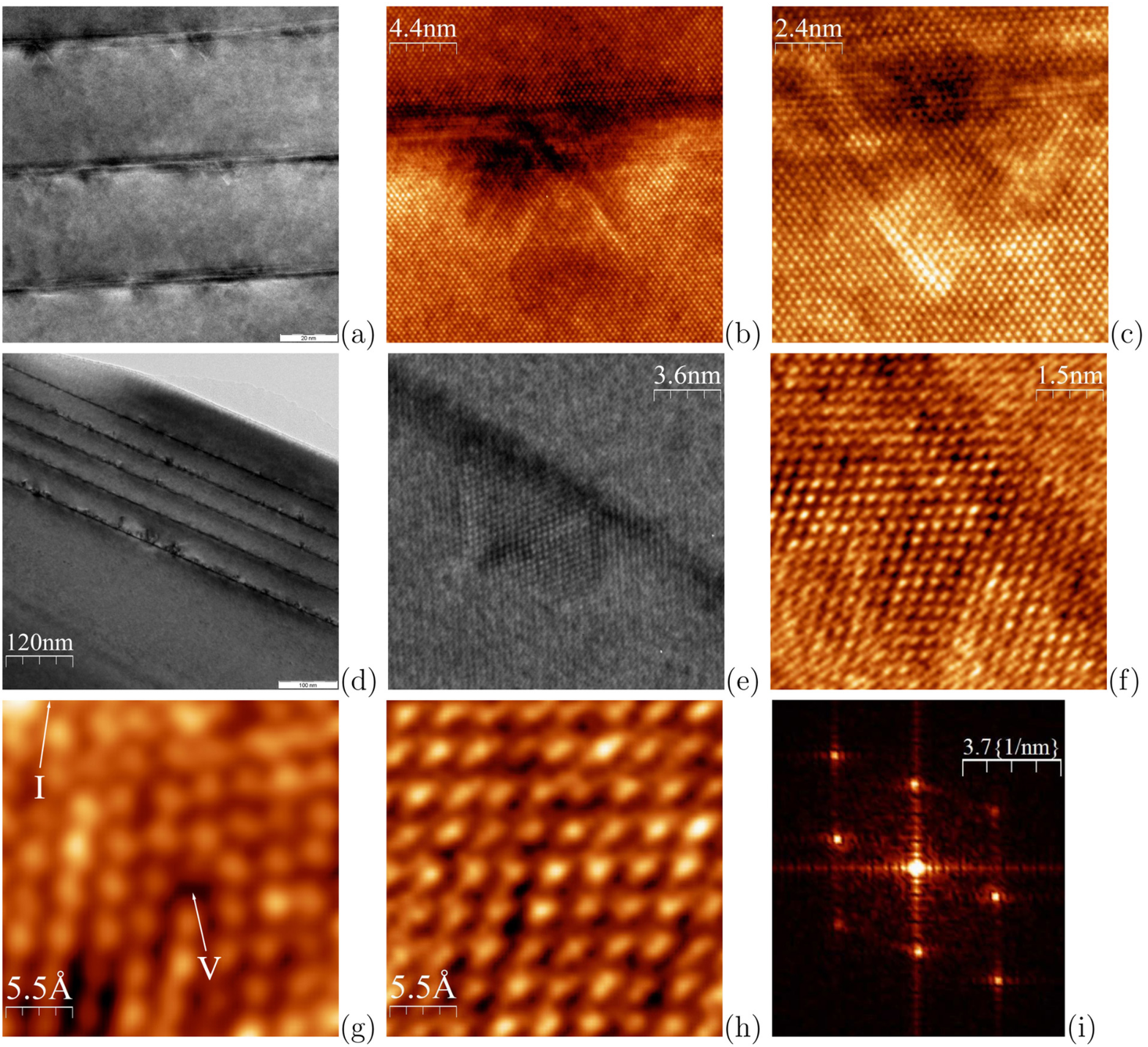}
\caption{\label{fig:TEM-9-10A}
}
\end{figure*}
\subsection*{Figure~\ref{fig:TEM-9-10A} - 
TEM data for the five-layer Ge/Si heterostructures, {\textit{T$_{\mathrm{gr}}$}}\,= 360\textcelsius:
}
(a) to (c) {\textit{h$_{\mathrm{Ge}}$}}\,= 9\,\r{A};
(d) to (i) {\textit{h$_{\mathrm{Ge}}$}}\,= 10\,\r{A};
(a)   domains of tensile strain in Si over Ge clusters are observed more or less distinctly near most clusters, but not around all; the surface is down; the mark equals 20 nm; 
(b), (c) zoom in two strained domains, no extended defects are observed; 
(d) strained domains are more pronounced, the strain is well recognized even under some clusters;
(e) a magnified image of a strained domain; a strained lattice is   well contrasted with the normal one;
(f) zoom in the dilated lattice, a perfectly ordered lattice is observed;
(g) a Si domain next to the Ge/Si interface near the cluster apex, a vacancy (`V') and disordered lattice (upper right corner) are revealed; letter `I' indicates the direction to the interface along {{\textless}11$\overline{1}$\textgreater};
(h) the same as in (g) but some farther from the interface, the lattice is perfect;
(i) the Fourier transform of an image obtained from a strained domain demonstrates an enhanced lattice parameter (the strain varies from domain to domain, the estimated lattice period in the [001] direction sometimes reaches $\sim$\,5.6\,\AA).
\pb

\clearpage
\begin{figure*}[h]
\includegraphics[scale=1]{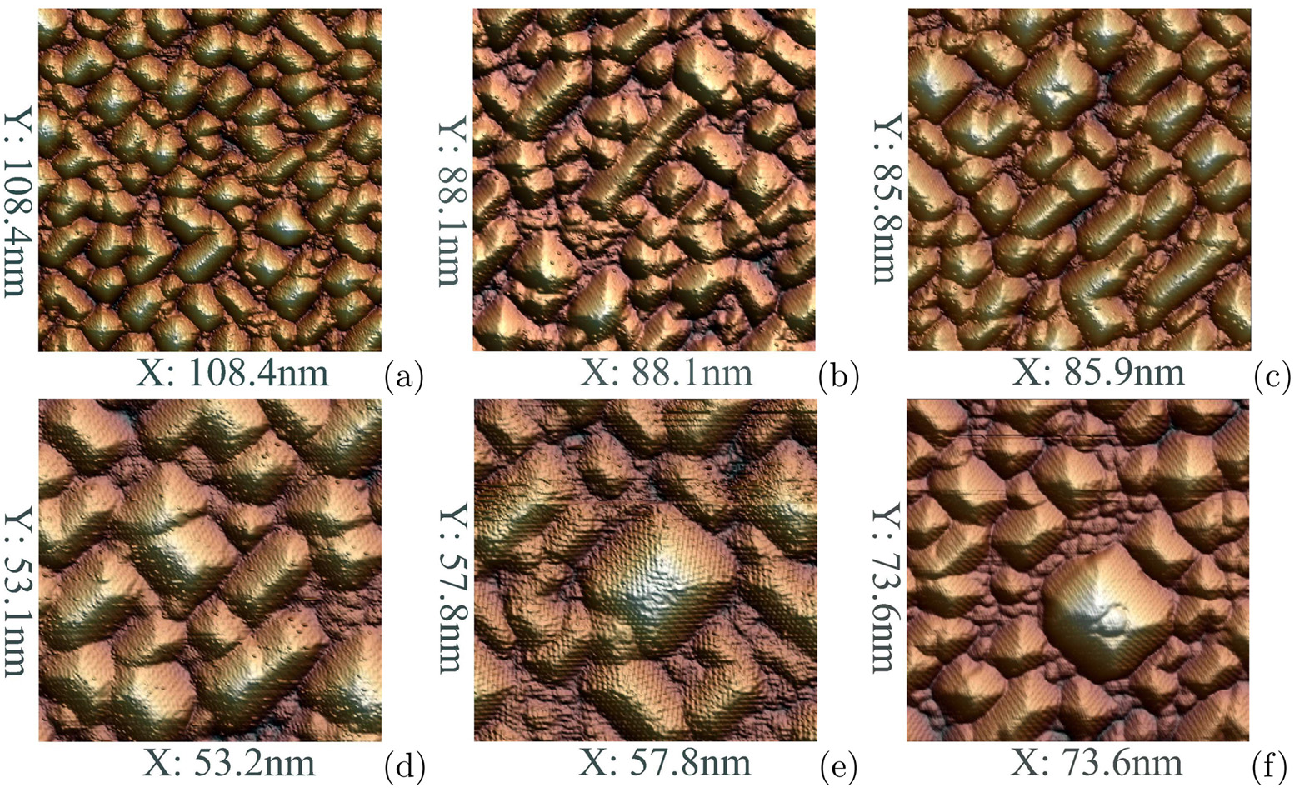}
\caption{\label{fig:STM-9A}
}
\end{figure*}
\subsection*{Figure~\ref{fig:STM-9A} - 
STM images of Ge/Si(001), {\textit{h$_{\mathrm{Ge}}$}}\,= 9\,\r{A}, {\textit{T$_{\mathrm{gr}}$}}\,= 360\textcelsius:
}
(a) to  
(d) array top views  with different magnifications;
(e) a large cluster in the array,  $\sim$\,2,5 nm high;
(f) a huge cluster (\textgreater\,3,5 nm high) interpreted as an array defect.
\pb

\clearpage
\begin{figure*}[h]
\includegraphics[scale=1]{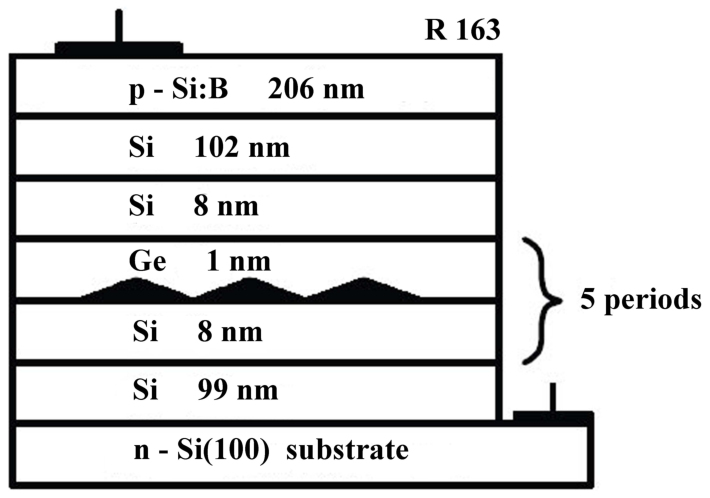}(a)\\
\includegraphics[scale=1]{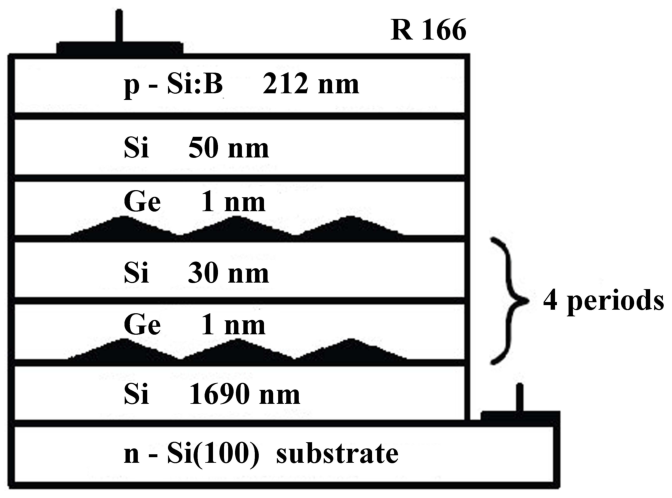}(b)
\caption{\label{fig:p-i-n_Schematics}  
}
\end{figure*}
\subsection*{Figure~\ref{fig:p-i-n_Schematics} - 
Schematics of the {\textit{p--i--n}}-structures:
}
(a) R\,163,  
(b) R\,166. 
\pb

\clearpage
\begin{figure*}
\Large
\includegraphics[scale=1]{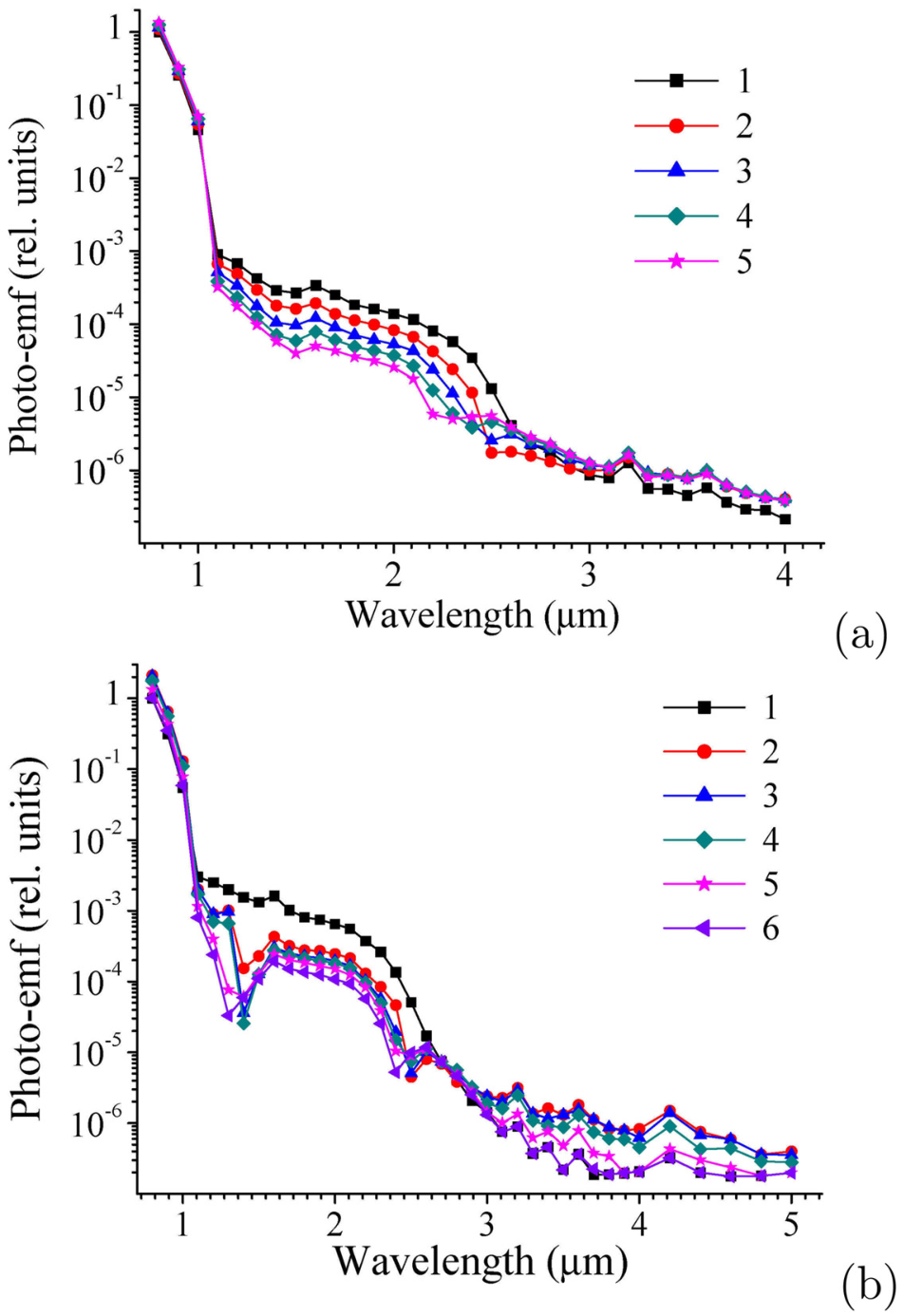}
\caption{\label{fig:r163_r166}
}
\end{figure*}
\subsection*{Figure~\ref{fig:r163_r166} -
Photo-emf spectra of the {\textit{p--i--n}} structures:
}
(a)~R\,163:
(1)~without bias lighting; 
(2)--(5) under bias lighting (Ge filter):
(2)~$W=0.25$\,mW/cm$^2$;
(3)~$W=0.77$\,mW/cm$^2$;
(4)~$W=1.5$\,mW/cm$^2$;
(5)~$W=2.16$\,mW/cm$^2$;
(b)~R\,166:
(1)~without bias lighting; 
(2)--(6) under bias lighting (Si filter):
(2)~$W=0.63$\,mW/cm$^2$;
(3)~$W=3.3$\,mW/cm$^2$;
(4)~$W=5.3$\,mW/cm$^2$;
(5)~$W=12$\,mW/cm$^2$;
(6)~$W=17.5$\,mW/cm$^2$.
\pb

\clearpage
\begin{figure*}
\includegraphics[scale=1.5]{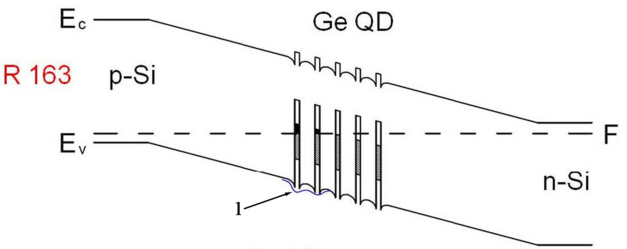}(a)\\
\includegraphics[scale=1.5]{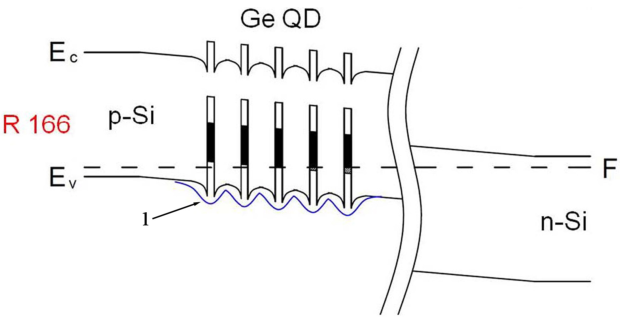}(b)
\caption{\label{fig:bands}
}
\end{figure*}
\subsection*{Figure~\ref{fig:bands} -
Schematics of band structures of {\textit{p--i--n}}-diodes:
}
(a) R\,163, 
(b) R\,166;  
figure `1' indicates potential barriers for holes in the valence band.
\pb

\clearpage
\begin{figure*}
\includegraphics[scale=1.5]{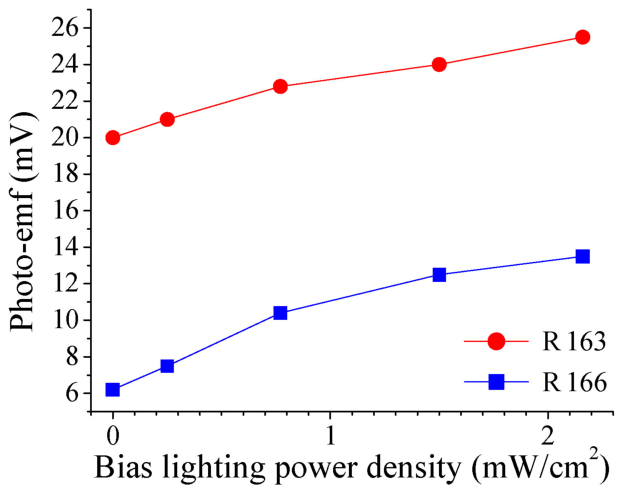}
\caption{\label{fig:bias}
 }
\end{figure*}
\subsection*{Figure~\ref{fig:bias} -
Dependence of photo-emf response of the R\,163 and R\,166 {\textit{p--i--n}}-structures on bias lighting power density.
}

\clearpage
\begin{figure*}[h]
\includegraphics[scale=.7]{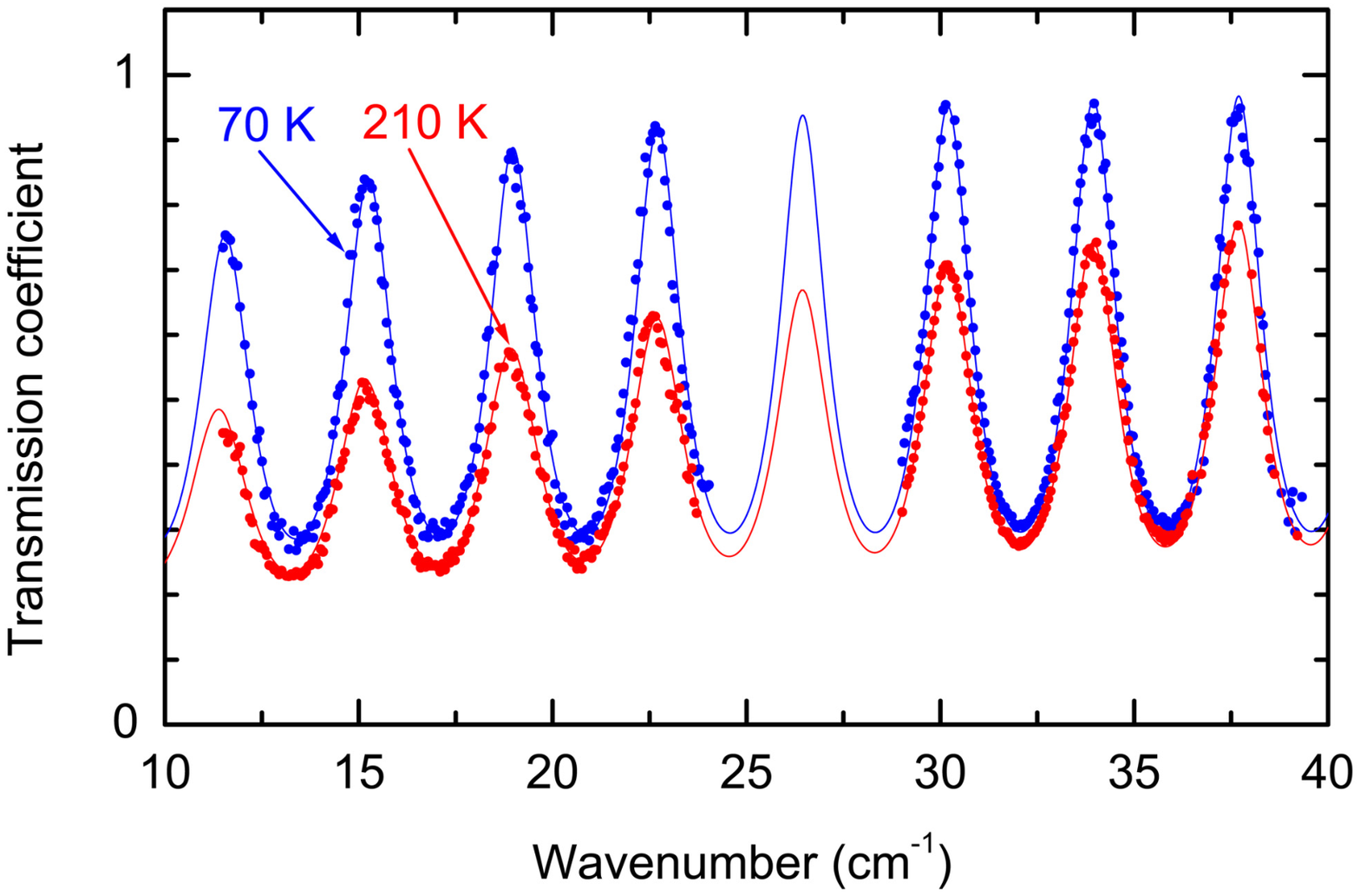}
\caption{\label{fig:THz-fig1}
}
\end{figure*}
\subsection*{Figure~\ref{fig:THz-fig1} - 
Spectra of transmission coefficient of a silicon substrate (a commercial wafer, $\rho = 12\,\Omega$\,cm), measured at two temperatures using two different BWO working in spectral ranges from 11 cm$^{-1}$ to 24 cm$^{-1}$ and from 29 cm$^{-1}$ to 39 cm$^{-1}$: 
}
Dots show the measurement results, lines are least-square fits based on the Drude conductivity model, as described in the text. 
\pb

\clearpage
\begin{figure*}[h]
\includegraphics[scale=.7]{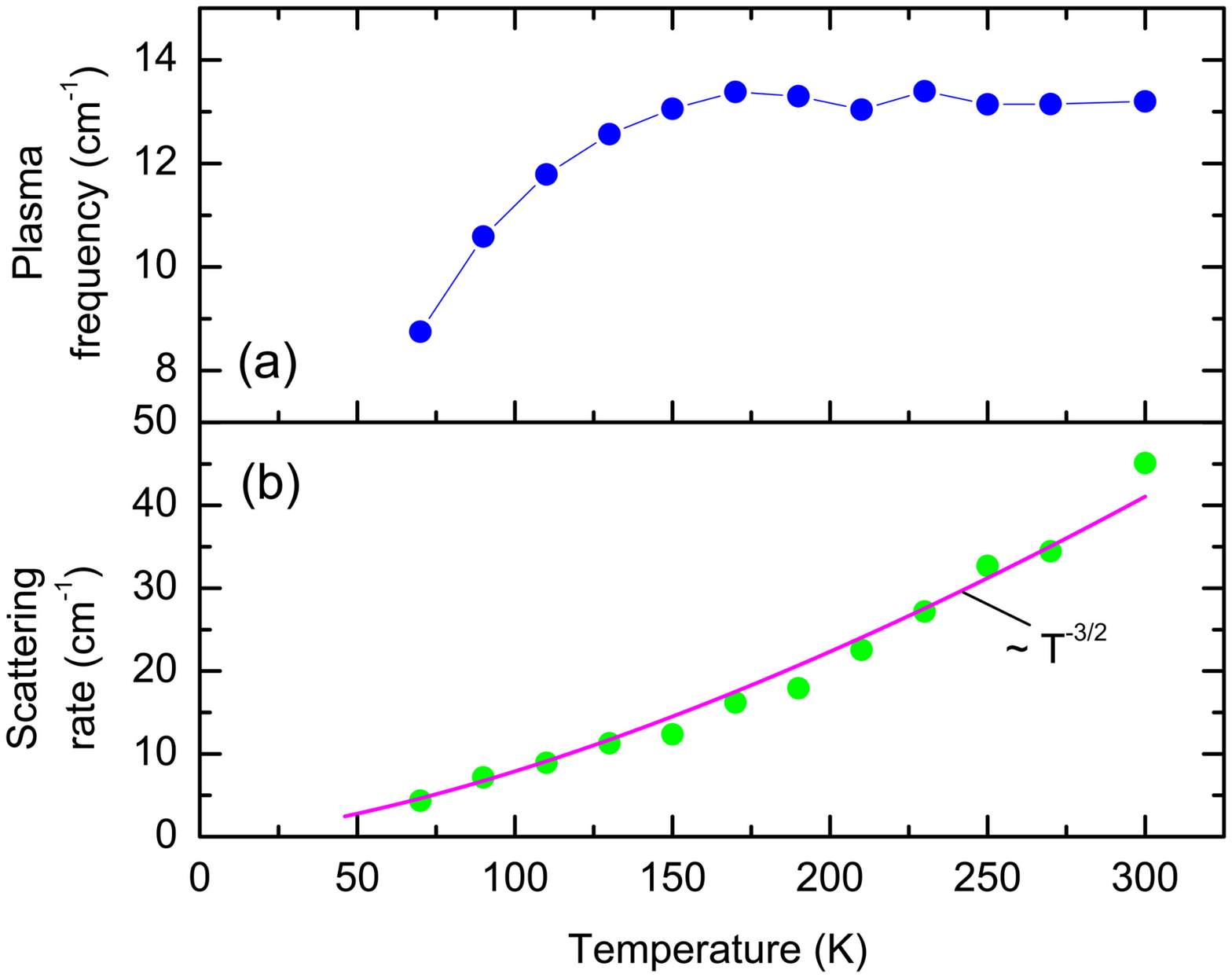}
\caption{\label{fig:THz-fig2}
}
\end{figure*}
\subsection*{Figure~\ref{fig:THz-fig2} - 
Temperature dependences of the silicon substrate parameters obtained by fitting the transmission coefficient spectra as shown in Figure~\ref{fig:THz-fig1} and described in the text: 
}
(a) plasma frequency of charge carriers  and 
(b) scattering rate. 
Solid line in (b) shows the $T^{-3/2}$ behavior.
\pb

\clearpage
\begin{figure*}[h]
\includegraphics[scale=.5]{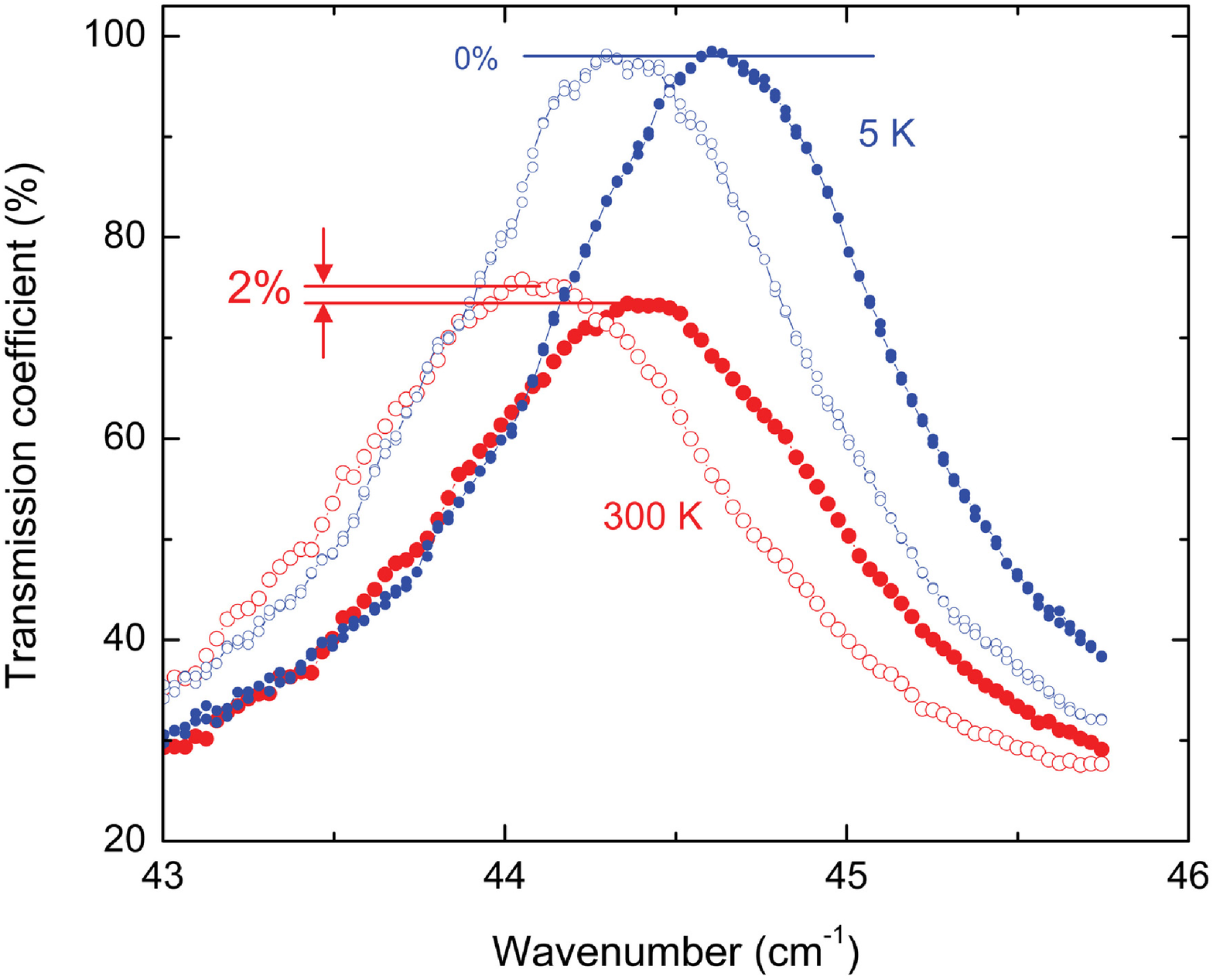}
\caption{\label{fig:THz-fig3}
}
\end{figure*}
\subsection*{Figure~\ref{fig:THz-fig3} - 
Spectra of transmission coefficient of Ge/Si heterostructure on Si substrate (solid symbols) and of bare substrate (open symbols) measured at two different temperatures: 
}
Horizontal lines show the difference in peak transmissivity that is observed at 300 K and disappears at $\sim$\,170 K. The peaks positions are shifted due to slight difference in the Si substrate thickness.
\pb

\clearpage
\begin{figure*}[h]
\includegraphics[scale=.7]{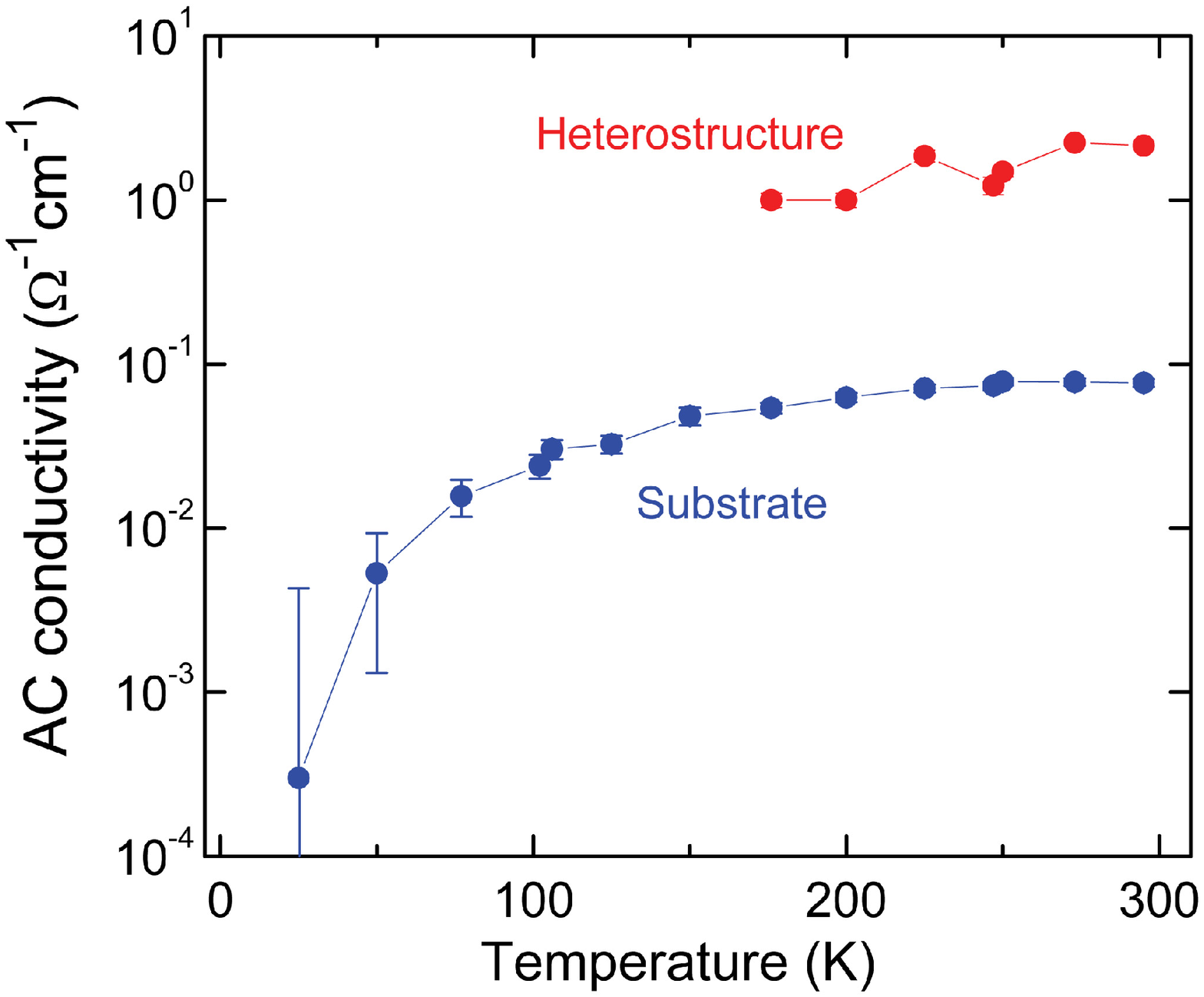}
\caption{\label{fig:THz-fig4}
}
\end{figure*}
\subsection*{Figure~\ref{fig:THz-fig4} - 
Temperature dependences of dynamical conductivity of Ge/Si heterostructure and of Si substrate:
}
Frequency is around 1 THz.
\pb

\clearpage
\begin{figure*}[h]
\includegraphics[scale=.5]{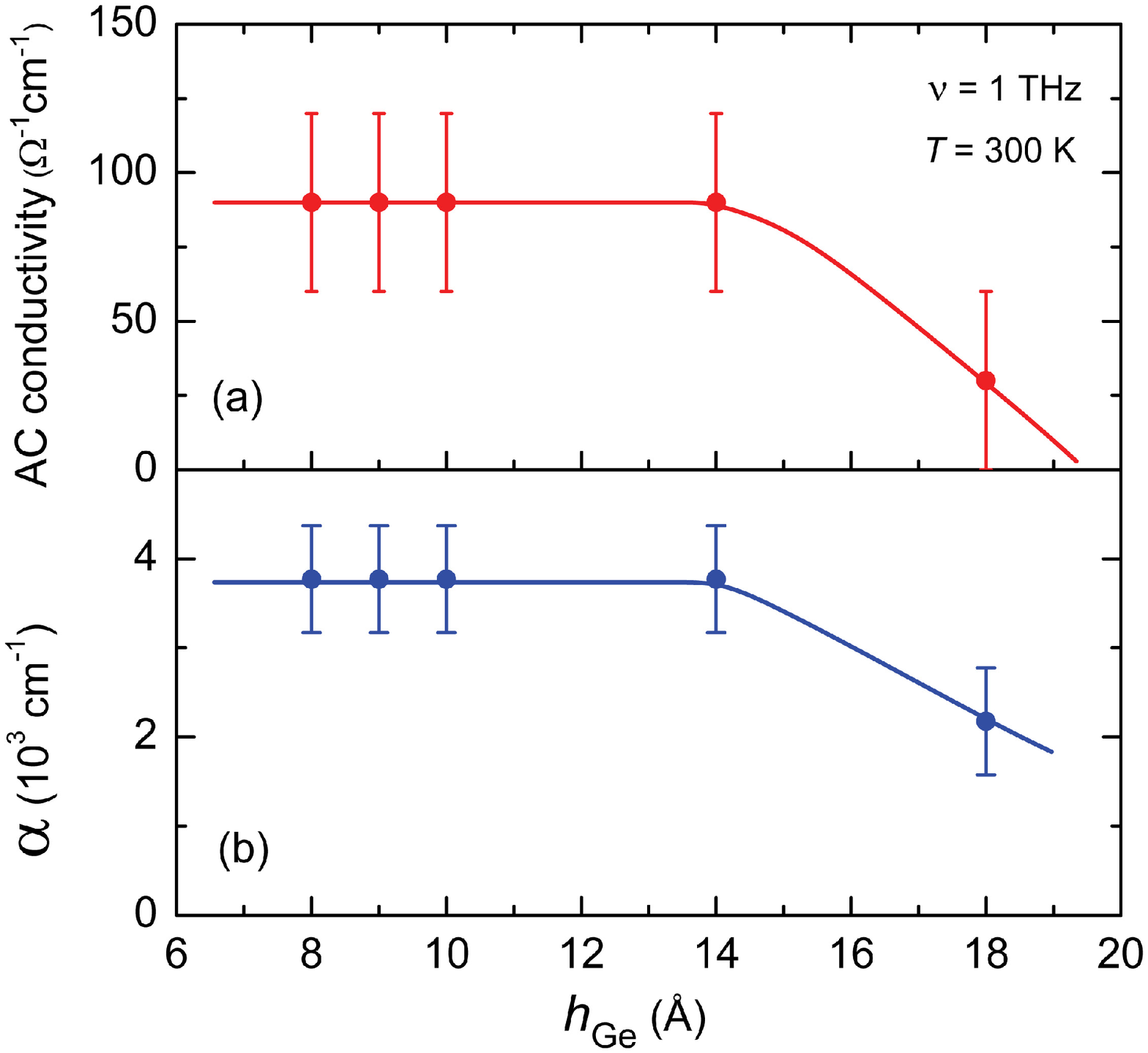}
\caption{\label{fig:THz-fig5}
}
\end{figure*}
\subsection*{Figure~\ref{fig:THz-fig5} - 
Terahertz conductivity and absorption coefficient of Ge/Si heterostructure with Ge quantum dots versus Ge coverage:
}
(a) terahertz conductivity, 
(b) absorption coefficient; 
lines are guides to the eye.
\pb

\end{bmcformat}
\end{document}